\documentclass[10pt,twocolumn,letterpaper]{article}

\usepackage[pagenumbers]{cvpr} 

\usepackage{multirow}   
\usepackage{algorithm}
\usepackage{algpseudocode} 
\usepackage{colortbl}
\usepackage{amsmath}

\definecolor{cvprblue}{rgb}{0.21,0.49,0.74}
\usepackage[pagebackref,breaklinks,colorlinks,allcolors=cvprblue]{hyperref}

\def\paperID{13732} 
\def\confName{CVPR}
\def\confYear{2025}

\title{vesselFM: A Foundation Model for Universal 3D Blood Vessel Segmentation}

\author{
Bastian Wittmann\textsuperscript{1}\quad 
Yannick Wattenberg\textsuperscript{2}\\
Tamaz Amiranashvili\textsuperscript{1,3}\quad 
Suprosanna Shit\textsuperscript{1}\quad 
Bjoern Menze\textsuperscript{1}\\
{\normalsize \textsuperscript{1}University of Zurich}\quad 
{\normalsize \textsuperscript{2}ETH Zurich}\quad 
{\normalsize \textsuperscript{3}Technical University of Munich}\\
{\tt\small \{bastian.wittmann, bjoern.menze\}@uzh.ch}
}

\begin{document}
\maketitle

\begin{abstract}
Segmenting 3D blood vessels is a critical yet challenging task in medical image analysis. This is due to significant imaging modality-specific variations in artifacts, vascular patterns and scales, signal-to-noise ratios, and background tissues. These variations, along with domain gaps arising from varying imaging protocols, limit the generalization of existing supervised learning-based methods, requiring tedious voxel-level annotations for each dataset separately. While foundation models promise to alleviate this limitation, they typically fail to generalize to the task of blood vessel segmentation, posing a unique, complex problem. In this work, we present vesselFM, a foundation model designed specifically for the broad task of 3D blood vessel segmentation. Unlike previous models, vesselFM can effortlessly generalize to unseen domains. To achieve zero-shot generalization, we train vesselFM on three heterogeneous data sources: a large, curated annotated dataset, data generated by a domain randomization scheme, and data sampled from a flow matching-based generative model. Extensive evaluations show that vesselFM outperforms state-of-the-art medical image segmentation foundation models across four (pre-)clinically relevant imaging modalities in zero-, one-, and few-shot scenarios, therefore providing a universal solution for 3D blood vessel segmentation.
\end{abstract}    
\section{Introduction and Motivation}\label{sec:intro}
Blood vessel segmentation represents a (pre-)clinically relevant task in (bio)medical image analysis as it plays a vital role in analyzing, diagnosing, and treating various vascular disorders, such as stroke~\cite{deshpande2021automatic}, cerebral aneurysms~\cite{nishi2024deep}, viral pneumonia~\cite{poletti2022automated}, coronary artery disease~\cite{xian2020main}, and Alzheimer's~\cite{walek2023near}.
Despite advances in medical image analysis, accurate and robust segmentation of fully-connected vasculature in task-specific imaging modalities still remains a challenging problem, especially in 3D. This is primarily due to the complexity introduced by intricate minuscule vascular geometries, as well as significant domain gaps caused by imaging modality and protocol-specific variations in signal-to-noise ratios, vascular patterns and scales, imaging artifacts, and background tissues. These variations typically prevent supervised deep learning-based methods from generalizing to unseen 3D blood vessel domains~\cite{wittmann2024simulation}. Consequently, researchers and clinicians frequently find themselves forced to default to the labor-intensive process of acquiring manual, voxel-level consistent annotations from scratch for analyzing vascular images at hand.

\begin{figure}[t]
\centerline{\includegraphics[width=0.9\linewidth]{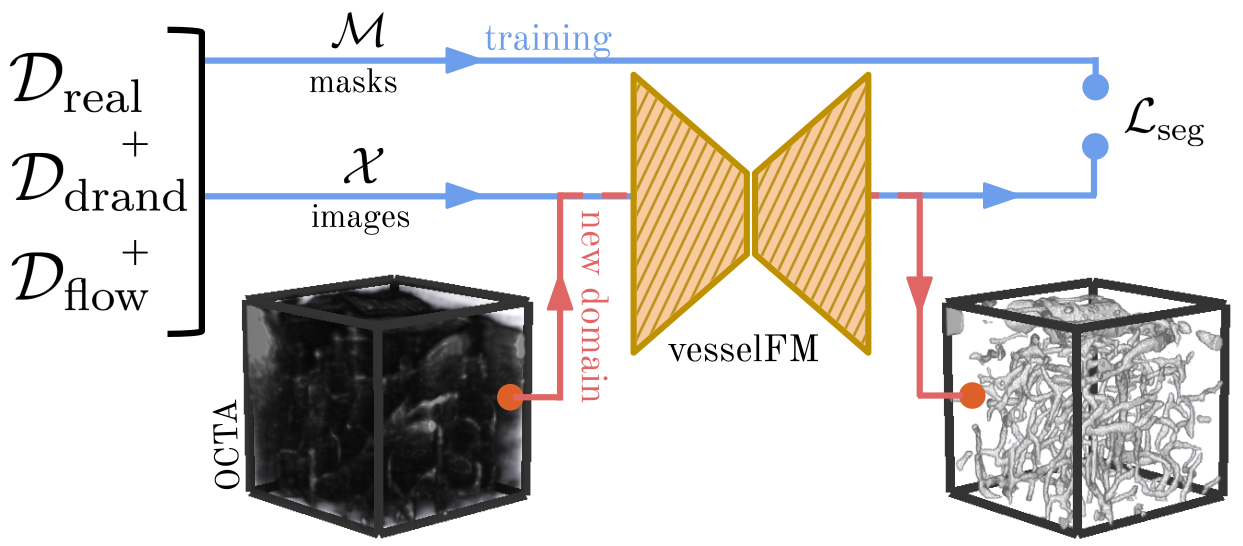}}
\caption{
VesselFM is trained in a supervised manner on image-mask pairs from three heterogeneous data sources ($\mathcal{D}_\text{real}$, $\mathcal{D}_\text{drand}$, and $\mathcal{D}_\text{flow}$) and subsequently applied in a \textit{zero}-, \textit{one}-, or \textit{few}-shot fashion to new, unseen 3D blood vessel domains.
}
\label{fig:overview}
\end{figure}

\begin{figure*}[th]
\centerline{\includegraphics[width=\linewidth]{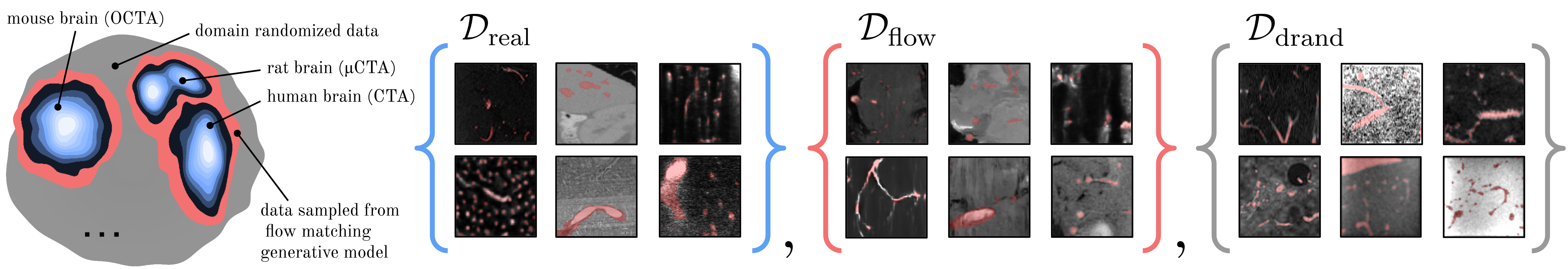}}
\caption{
Schematic distributions of our three data sources $\mathcal{D}_\text{real}$ (shades of blue), $\mathcal{D}_\text{flow}$ (red), and $\mathcal{D}_\text{drand}$ (gray). While we aim to comprehensively cover the general domain of 3D vascular images with $\mathcal{D}_\text{drand}$, $\mathcal{D}_\text{flow}$ effectively broadens the distributions of $\mathcal{D}_\text{real}$. Note that segmentation masks are shown in translucent red in the exemplary images.
}
\label{fig:domain_overview}
\end{figure*}

Foundation models for image segmentation pre-trained on large-scale datasets have recently emerged as tools that can effortlessly generalize to unseen data distributions~\cite{kirillov2023segment}. Although segmentation foundation models also established themselves in the medical field~\cite{ma2024segment,zhu2024medical,wang2024sammed3d}, they typically fail to overcome the unique challenges posed by the characteristics of the vascular network.
To address this limitation, we propose vesselFM, a \underline{F}oundation \underline{M}odel precisely tailored to universal 3D blood vessel segmentation. We train vesselFM in a supervised manner on image-mask pairs from three heterogeneous data sources (see Fig.~\ref{fig:overview}). To this end, we first curate $\mathcal{D}_\text{real}$, which represents, to the best of our knowledge, the largest annotated dataset for 3D blood vessel segmentation, covering a broad range of imaging modalities from various anatomical regions of different organisms. Second, we supplement $\mathcal{D}_\text{real}$ with two synthetic data sources, $\mathcal{D}_\text{drand}$ and $\mathcal{D}_\text{flow}$. In particular, we aim to comprehensively cover the general domain of 3D vascular images by adopting strategies from domain randomization, while we aim to additionally broaden data distributions included in $\mathcal{D}_\text{real}$ by sampling from a mask- and class-conditioned flow matching-based generative model (see Fig.~\ref{fig:domain_overview}). Constructing vesselFM on our proposed data sources results in robust features that enable strong generalization to unseen imaging domains, facilitating broad use. In extensive experiments, we demonstrate vesselFM's state-of-the-art performance on the tasks of \textit{zero}-, \textit{one}-, and \textit{few}-shot blood vessel segmentation across four (pre-)clinically relevant datasets.

\noindent Our contributions can be summarized as follows:
\begin{enumerate}
    \item We propose a universal foundation model for 3D blood vessel segmentation capable of \textit{zero}-shot generalization. By open-sourcing checkpoints and code, we aim to provide a foundation model that serves as a robust, out-of-the-box tool for researchers and clinicians alike.\footnote{\url{https://github.com/bwittmann/vesselFM}}
    
    \item $\mathcal{D}_\text{real}$: We curate the largest dataset for 3D blood vessel segmentation, consisting of carefully processed, real 3D vascular images with matching voxel-level annotations.
    
    \item $\mathcal{D}_\text{drand}$: We propose an elaborate domain randomization strategy tailored to 3D blood vessel segmentation.

    \item $\mathcal{D}_\text{flow}$: We introduce mask- and class-conditioned flow matching to 3D medical image generation, producing high-fidelity image-mask pairs that adhere to coherent anatomical constraints.
\end{enumerate}
\section{Related Works}\label{sec:related_works}
In this section, we discuss works closely related to vesselFM and elaborate on how vesselFM distinguished itself from existing literature.

\subsection{Foundation Models for Image Segmentation}
The advent of the Segment Anything Model (SAM)~\cite{kirillov2023segment} led to the creation of several SAM-like foundation models designed for medical image segmentation~\cite{ma2024segment,zohranyan2024dr,wang2024samocta}, even in 3D~\cite{wang2024sammed3d,wu2023medical,wan2024trisam,zhu2024medical}.
The general-purpose segmentation model SAM-Med3D~\cite{wang2024sammed3d}, \eg, is trained on a combination of 94 datasets, offers generalization across anatomical structures and imaging modalities, and claims \textit{zero}-shot transferability to unseen tasks. In contrast to SAM-Med3D, MedSAM-2~\cite{zhu2024medical} relies on the updated SAM 2~\cite{ravi2024sam} and follows the philosophy of treating 3D medical images as videos, resulting in state-of-the-art results while maintaining exceptional generalization across a wide variety of imaging modalities. VISTA3D~\cite{he2024vista3d}, on the other hand, is developed specifically for CT scans. VISTA3D segments 127 structures and lesions in highly variant CT scans, offering accurate out-of-the-box results and effortless adaptation to unseen structures.
In the realm of vessel segmentation, however, foundation models remain under-explored. Earlier works experimented with fine-tuning models pre-trained on vascular data~\cite{tetteh2020deepvesselnet, holroyd2023tube}, \textit{few}-shot learning~\cite{aktar2023vesselshot}, and SAM-like methods tailored to 2D OCTA images~\cite{wang2024samocta}, 3D vEM images~\cite{wan2024trisam}, and 2D X-ray images~\cite{zohranyan2024dr}. Unlike vesselFM, all above-mentioned vessel segmentation methods are either limited to specific imaging modalities and anatomical structures or have a significantly narrower scope.

\begin{table*}[ht]
\centering
\tiny
\caption{
Overview of $\mathcal{D}_\text{real}$, including selected dataset statistics. We estimate the mean shape over the $x$-, $y$-, and $z$-axis individually. The voxel size represents the spatial resolution at acquisition time, while the number of patches reflects the approximate amount of $\text{128}^\text{3}$ sub-volumes comprising $\mathcal{D}_\text{real}$. We further provide an estimate on label quality focusing on vessel connectivity and annotation precision and a brief overview of dataset-specific pre-processing steps, ensuring that datasets comply with our quality standards. Additional details are provided in Suppl.~\ref{suppl:preprocessing}.
Note that the first four datasets are exclusively used to evaluate vesselFM on unseen domains in our experiments.
}
\label{tab:datasets}
\begin{tabular}{c|l|c|l|l|l|l|l|l|l|l} 
\toprule
&Name & Class $c$ & Tissue Type & Imaging Modality & \# Images & Mean Shape & Voxel Size & \# Patches & Label Quality & Pre-Processing$^{**}$\\
\midrule
\multirow{4}{*}{\rotatebox[origin=c]{90}{\textit{Evaluation}}}
&SMILE-UHURA~\cite{chatterjee2024smile}& 1 & human brain & MRA & 14 & 640 $\times$ 482 $\times$ 163 & 0.30$\times$0.30$\times$0.30 mm & 335 & 9 & -\\
&BvEM~\cite{wan2024trisam}& 2 & mouse brain  & vEM & 1 & 3571 $\times$ 5145 $\times$ 2495 & 0.25$\times$0.25$\times$0.32 \textmu m & 21858 & 8 & r, c\\
&OCTA~\cite{wittmann2024simulation,glandorf2024bessel} & 3 & mouse brain & OCTA & 6 & 160 $\times$ 160 $\times$ 160 & 2.00$\times$2.00$\times$2.00 \textmu m & 11 & 10 & -\\
&MSD8~\cite{antonelli2022medical}& 4 & human liver & CT & 443 & 512$\times$ 512 $\times$ 70 & 0.80$\times$0.80$\times$5.00 mm & 2640 & 9 & c, ic\\
\midrule
&TubeTK~\cite{bullitt2005vessel}& 5 & human brain & MRA & 42 & 896 $\times$ 896 $\times$ 256 & 0.50$\times$0.50$\times$0.80 mm & 4116 & 8 & r, mp\\[0.5mm]
&\multirow{2}{*}{tUbeNet~\cite{holroyd2023tube}} & 6& mouse liver & HREM MRI  & 1 & 400 $\times$ 400 $\times$ 89 & 0.90$\times$0.90$\times$5.00 mm & 6 & 6 & -\\
&& 7 & mouse brain & two-photon microscopy & 1 & 500$\times$ 500 $\times$ 356 & 0.20$\times$0.46$\times$5.20 \textmu m & 42 & 7 & r, mp\\[0.5mm]
&\multirow{2}{*}{TopCoW~\cite{yang2023benchmarking}} & 8& human brain & CTA & 90 & 334 $\times$ 451 $\times$ 128 & 0.45$\times$0.45$\times$0.13 mm & 863 & 8 & r\\
&& 9 & human brain & MRA & 90 & 406 $\times$ 522 $\times$ 128 & 0.30$\times$0.30$\times$0.60 mm & 1179 & 8 & r\\[0.5mm]
&\multirow{2}{*}{VesSAP~\cite{todorov2020machine}} & 10 & mouse brain & light-sheet microscopy (EB)$^{*}$ & 19 & 500$\times$ 500 $\times$ 50 & 2.83$\times$2.83$\times$4.99 \textmu m & 113 & 7 & ic\\
&& 11 & mouse brain & light-sheet microscopy (WGA)$^{*}$ & 19 & 500$\times$ 500 $\times$ 50 & 2.83$\times$2.83$\times$4.99 \textmu m & 113 & 7 & ic\\[0.5mm]
&\multirow{2}{*}{DeepVesselNet~\cite{tetteh2020deepvesselnet}} & 12 & human brain & MRA & 40 & 544$\times$ 514 $\times$ 132 & 0.31$\times$0.31$\times$0.60 mm & 726 & 7 & r\\
&& 13 & rat brain & \textmu CTA & 4 &  256 $\times$ 256 $\times$ 256 & 0.70$\times$0.70$\times$0.70 mm & 32 & 7 & -\\[0.5mm]
&HR-Kidney~\cite{kuo2023terabyte} &14& mouse kidney & X-ray & 1 & 4608$\times$ 4608 $\times$ 7168 & 1.60$\times$1.60$\times$1.60 \textmu m & 72576 & 6 & mp\\
&3D-IRCADb-01~\cite{soler20103d} &15& human liver  & CT & 20 & 512$\times$ 512 $\times$ 141 & 0.57$\times$0.57$\times$1.60 mm & 352 & 6 & c, mp\\
&DeepVess~\cite{haft2019deep} &16& mouse brain & multi-photon microscopy & 1 & 256$\times$256$\times$200 & 1.00$\times$1.00$\times$1.70 \textmu m & 6 & 8 & -\\
&CSD~\cite{chen2022attention,ixidataset} &17& human brain & MRA & 45 &  1024$\times$ 1024$\times$ 92 & 0.26$\times$0.26$\times$0.80 mm & 2070 & 7 & -\\[0.5mm]
&\multirow{3}{*}{VesselExpress~\cite{spangenberg2023rapid}}& 18 & mouse brain & light-sheet microscopy & 4 & 2000$\times$2000$\times$501 & 2.00$\times$2.00$\times$8.00 \textmu m & 3822 & 6 & -\\
&& 19 & mouse heart & light-sheet microscopy & 3 &  250 $\times$ 250 $\times$ 222  & 2.00$\times$2.00$\times$8.00 \textmu m & 19 & 6 & -\\
&& 20 & mouse bladder & light-sheet microscopy & 10 &  300$\times$300$\times$101 & 2.00$\times$2.00$\times$8.00 \textmu m & 43 & 6 & -\\[0.5mm]
&MiniVess~\cite{poon2023dataset}& 21 & mouse brain & two-photon microscopy & 70 & 512$\times$512$\times$43 & 0.70$\times$0.70$\times$5.00 \textmu m & 380 & 7 & -\\
&HiP-CT~\cite{yagis2023deep}& 22 & human kidney & CT & 3 & 1350$\times$ 1311 $\times$ 1844 & 2.50$\times$2.50$\times$2.50 \textmu m & 4225 & 8 & -\\
&LS~\cite{binder2024leptomeningeal}& 23 & mouse brain & light-sheet microscopy & 1 & 175$\times$170$\times$200 & 6.00$\times$6.00$\times$6.00 \textmu m & 2 & 9 & ic\\
\bottomrule

\multicolumn{10}{l}{\rule{0pt}{2mm} \tiny{$^{*}$different dyes used for staining: wheat germ agglutinin (WGA) and Evans blue (EB);} \tiny{$^{**}$r: resampled, c: cropped, mp: mask post-processed (\eg, smoothed or multi-class labels to binary), ic: intensities clipped}}\\
\end{tabular}
\end{table*}

\subsection{Synthetic Medical Image Generation}
In the medical domain, synthetic data is often used to address data scarcity or enhance data diversity. 
In this context, deep generative models, with diffusion models at the forefront, have emerged as a powerful technique for producing vast amounts of high-fidelity synthetic data~\cite{hamamci2023generatect, friedrich2024deep}. However, to leverage data generated by diffusion models for the task of segmentation, precisely matching image-mask pairs are required. Tackling this challenge, Med-DDPM~\cite{dorjsembe2024conditional} and SegGuidedDiff~\cite{konz2024anatomically} integrate semantic conditioning via channel-wise concatenation of the segmentation mask to the model input, resulting in image-mask pairs following coherent anatomical constraints. While Med-DDPM is tailored to 3D brain imaging synthesis, SegGuidedDiff experiments with 2D breast MRI and abdominal CT generation.
The concept of domain randomization~\cite{tobin2017domain} represents another promising technique in which fore- and background intensity values and morphological features are randomized in a semi-controlled manner to generate versatile synthetic image-mask pairs that can be utilized to train generalist segmentation models resilient to domain shifts. Billot \etal~\cite{billot2023synthseg} were the first to adopt a domain randomization strategy for medical image segmentation by proposing SynthSeg, a model capable of segmenting brain MRI scans of varied resolution and contrast. AnyStar~\cite{dey2024anystar} extends SynthSeg's concept to 3D instance segmentation of star-convex shapes such as nuclei, nodules, or metastases.
In this work, we leverage both deep generative models and domain randomization strategies to enrich our data sources. Specifically, we extend Med-DDPM by introducing the concept of flow matching~\cite{lipman2023flow,liu2022flow}, which has shown to be superior to diffusion~\cite{esser2024scaling,ma2024sit}, to anatomically controllable vascular image generation, and adapt the concept of domain randomization for 3D blood vessel segmentation.

\section{Data Source Generation}\label{sec:methods}
We train vesselFM on three heterogeneous data sources: 1) diverse real data ($\mathcal{D}_\text{real}$), 2) domain randomized data ($\mathcal{D}_\text{drand}$), and 3) data sampled from a flow matching-based generative model ($\mathcal{D}_\text{flow}$). Below, we detail each of these three data sources.

\begin{figure}[ht]
\centerline{\includegraphics[width=0.7\linewidth]{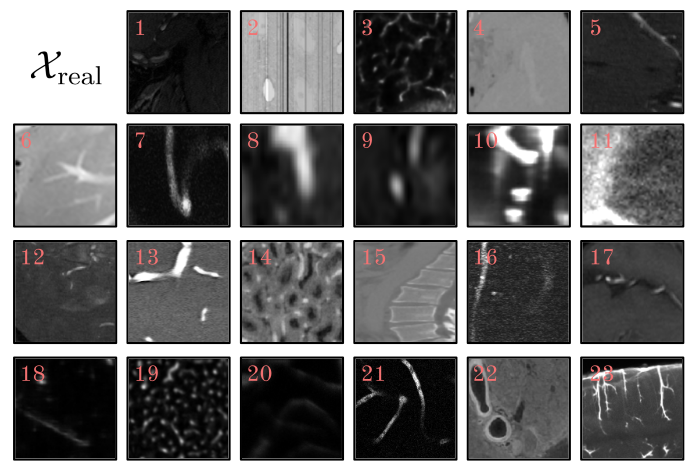}}
\caption{
Slices of images $\mathcal{X}_\text{real}$ from $\mathcal{D}_\text{real}$. $\mathcal{D}_\text{real}$ contains vascular images of shape $\text{128}^\text{3}$ with matching voxel-level annotations collected from 23 datasets (classes are indicated in red) of diverse imaging modalities, depicting a wide range of anatomical regions.
}
\label{fig:method_real}
\end{figure}

\begin{figure*}[t]
\centerline{\includegraphics[width=\linewidth]{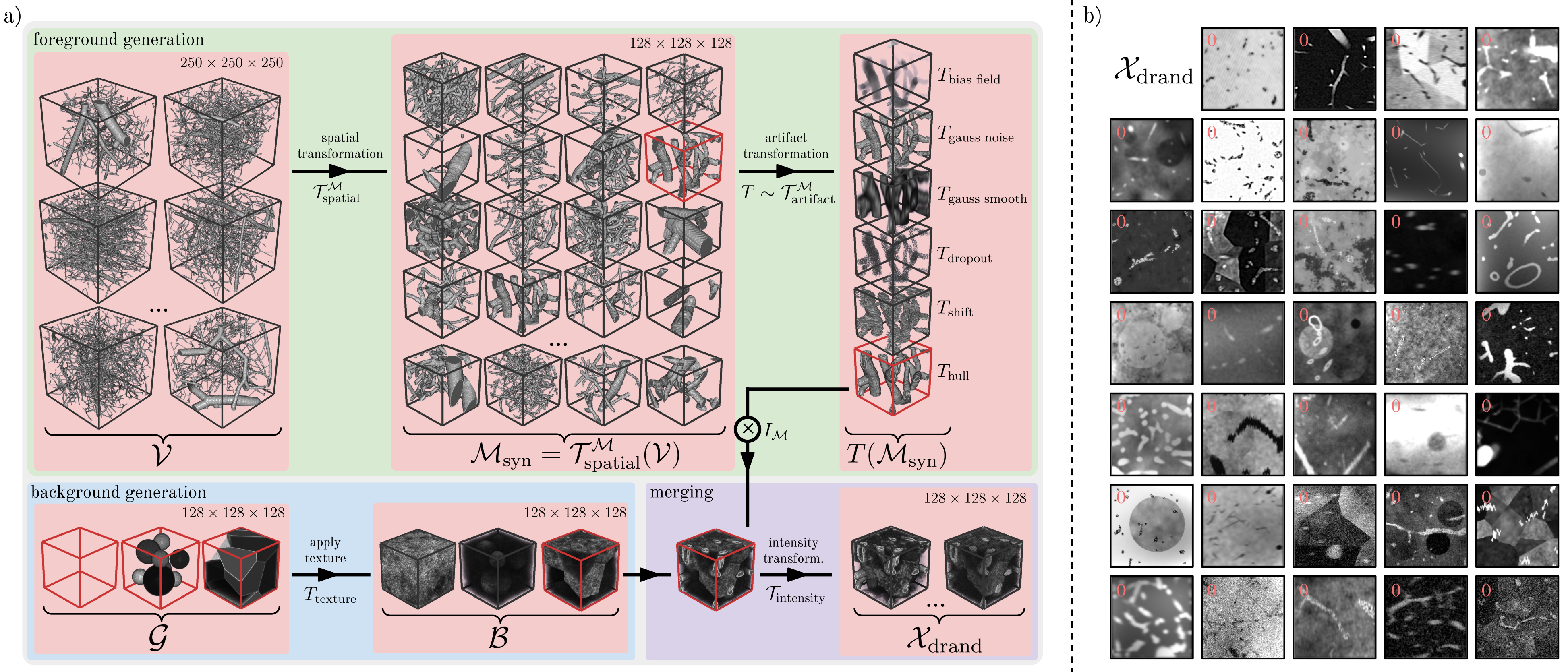}}
\caption{
a) Schematic overview of our domain randomized generative pipeline used to generate $\mathcal{D}_\text{drand}$ = $\{\mathcal{X}_\text{drand}, \mathcal{M}_\text{syn}\}$. We specifically highlight its three main components: foreground generation, background generation, and merging. Note that we indicate instances forwarded to the subsequent step in the color red for illustration purposes.
b) Slices of exemplary images $\mathcal{X}_\text{drand}$, categorized as $c = 0$. The wide variety of generated, highly diverse images showcases the effectiveness of our proposed domain randomization strategy.
}
\label{fig:method_drand}
\end{figure*}

\subsection{\texorpdfstring{$\mathcal{D}_\text{real}$:}{} Diverse Real Data}
The development process of generalist foundation models necessitates large-scale, diverse real datasets~\cite{ma2024segment,wang2024sammed3d,butoi2023universeg}. To this end, we curate $\mathcal{D}_\text{real}$ = $\{\mathcal{X}_\text{real}, \mathcal{M}_\text{real}\}$, encompassing real images $\mathcal{X}_\text{real}$ and their corresponding annotated segmentation masks $\mathcal{M}_\text{real}$ (see Table~\ref{tab:datasets} for dataset overview and statistics; Fig.~\ref{fig:method_real} for exemplary images). 
$\mathcal{D}_\text{real}$ comprises more than 115,000 3D patches of shape $\text{128}^\text{3}$ curated from 17 annotated sources, which we further separate into 23 datasets based on tissue types, imaging modalities, and protocols. For ease of reference, each dataset in $\mathcal{D}_\text{real}$ is indexed by a unique class $c \in \mathcal{C} = \{1, ..., 23\}$ (see Table~\ref{tab:datasets}, 2nd column). Importantly, $\mathcal{D}_\text{real}$ covers a broad array of clinically (\eg, MRA, CTA, and X-ray) and pre-clinically relevant (\eg, vEM, \textmu CTA, and two-photon microscopy) imaging modalities. It integrates data from several anatomical regions (\eg, brain, kidney, and liver) in various organisms (\eg, human, mouse, and rat), thus providing an expansive spectrum of blood vessel patterns of varying structural and functional properties. Further, we deliberately include datasets of the same imaging modalities to bridge domain gaps in, \eg, scale and contrast, induced by high variability in dataset-specific imaging protocols. We pay special attention to solely including datasets that adhere to a high standard in label quality. 
To curate $\mathcal{D}_\text{real}$, we pre-process each dataset and finally extract patches of our target shape ($\text{128}^\text{3}$) from the images and their corresponding labels. Details on pre-processing can be found in Suppl.~\ref{suppl:preprocessing}. To the best of our knowledge, $\mathcal{D}_\text{real}$ represents the largest real dataset for the task of 3D blood vessel segmentation to this date.

\subsection{\texorpdfstring{$\mathcal{D}_\text{drand}$:}{} Domain Randomization}
Inspired by recent works~\cite{billot2023synthseg,dey2024anystar}, we explore the use of domain randomization to generate a massive amount of matching image-mask pairs of semi-randomized style, categorizing them under class $c = 0$. In the following, we describe our proposed domain randomization strategy tailored to 3D blood vessels. An overview of our domain randomized generative pipeline used to create $\mathcal{D}_\text{drand}$ = $\{\mathcal{X}_\text{drand}, \mathcal{M}_\text{syn}\}$ is depicted in Fig.~\ref{fig:method_drand}. We detail its parametrization in Suppl.~\ref{suppl:param_drand}.

\paragraph{Foreground generation.}
To generate synthetic masks acting as foreground geometries, we utilize 1,137 vascular patches $\mathcal{V}$ of shape $\text{250}^{\text{3}}$ provided by Wittmann \etal~\cite{wittmann2024simulation}. These vascular patches, originating from graph representations of corrosion casts~\cite{walchli2021hierarchical}, accurately preserve both general angioarchitectural and morphological properties characteristic of 3D blood vessels with minimal artifacts. Therefore, $\mathcal{V}$ provides the perfect foundation by ensuring functional fidelity, a key requirement for generating realistic vascular images.
First, we process $\mathcal{V}$ by applying spatial transformations $\mathcal{T}_{\text{spatial}}^{\mathcal{M}}$. Specifically, we crop with random center positions to the target shape of $\text{128}^{\text{3}}$, followed by random flipping and rotation across all three axes. To ensure robustness against variations in blood vessel scale and density, we subsequently apply random dilation and random zooming. Additionally, we address variations in blood vessel curvature and tortuosity by employing random elastic deformation and binary smoothing. Applying $\mathcal{T}_{\text{spatial}}^{\mathcal{M}}$ results in a broad range of realistic vascular patterns (see Fig.~\ref{fig:method_drand}a). Throughout this work, we refer to $\mathcal{T}_{\text{spatial}}^{\mathcal{M}}(\mathcal{V})$ as the set $\mathcal{M}_\text{syn}$.
Next, we emulate a broad range of foreground artifacts present in real vascular images by concluding with carefully selected artifact transformations $\mathcal{T}_{\text{artifact}}^{\mathcal{M}}$ = $\{T_{\text{bias field}}$, $T_{\text{gauss noise}}$, $T_{\text{gauss smooth}}$, $T_{\text{dropout}}$, $T_{\text{shift}}$, $T_{\text{hull}}$, $T_{\text{identity}}\}$ (see Fig.~\ref{fig:method_drand}a). In contrast to $\mathcal{T}_{\text{spatial}}^{\mathcal{M}}$, which is applied consecutively, we sample a single artifact transformation $T$ from $\mathcal{T}_{\text{artifact}}^{\mathcal{M}}$ for each processed vascular patch.

\begin{figure*}[t]
\centerline{\includegraphics[width=\linewidth]{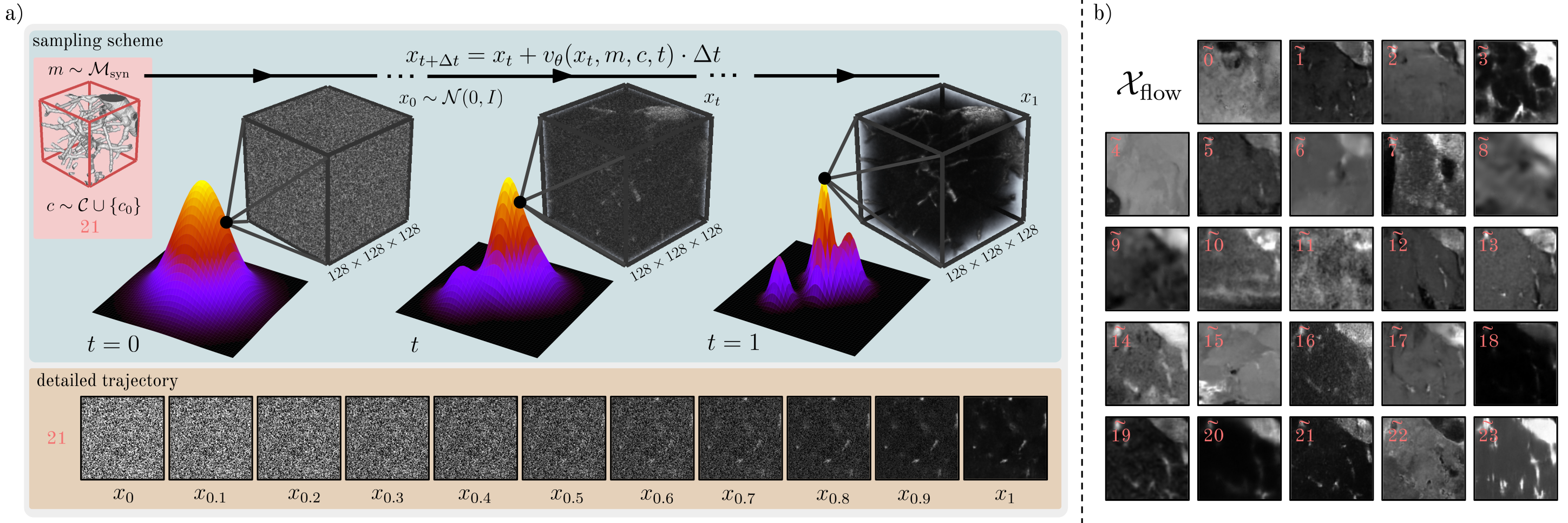}}
\caption{
a) Sampling of synthetic images $\mathcal{X}_\text{flow}$ via our mask- and class-conditioned flow matching-based generative model. We explicitly show our sampling scheme, mapping a sample $x_{0} \sim \mathcal{N}(0, I)$ to an exemplary sample $x_{1}$ of class $\Tilde{21}$. In addition, we present a more detailed trajectory, which is for improved visibility plotted in 2D.
b) Slices of exemplary images $\mathcal{X}_\text{flow}$, sampled from our generative model. Note that all of the depicted slices are conditioned on the same mask, and we solely vary the class. We would like to emphasize that our generative model is able to produce synthetic images almost indistinguishable from real images (compare with Fig.~\ref{fig:method_real}).
}
\label{fig:method_fm}
\end{figure*}

\paragraph{Background generation.}
Considering that the interplay of imaging techniques and protocols, background tissue compositions, and pathological conditions creates a broad spectrum of background intensity patterns, we model background images $\mathcal{B}$ containing various background geometries of diverse textures. Specifically, we incorporate three variants of background geometries $\mathcal{G}$ (see Fig.~\ref{fig:method_drand}a). 1) Spheres: we include non-overlapping spheres; 2) polyhedrons: we split the image into polyhedral regions using Voronoi partitioning~\cite{aurenhammer1991voronoi}; 3) none: we do not incorporate any background geometries. For background geometries and the background itself, we sample versatile Perlin noise patterns~\cite{perlin1985image,dey2024anystar} that accurately mimic textures characteristic of vascular images. Further, we also include plain background images consisting of a randomly selected intensity value drawn from $\mathcal{U}(0, 1)$ for enhanced diversity.

\paragraph{Fore- and background merging.}
Subsequently, we merge $T(\mathcal{M_\text{syn}})$ into the background images $\mathcal{B}$ via either voxel-wise addition/subtraction or by replacing background intensity values with mask intensity values. To assign intensities $I_{\mathcal{M}}$ to individual masks that separate them from their respectively matched background images, we estimate mean background intensities $I_{\mathcal{B}}^{\mu}$ and follow $I_{\mathcal{M}} \notin \left[I_{\mathcal{B}}^{\mu} - \delta, I_{\mathcal{B}}^{\mu} + \delta \right]$.
Lastly, we additionally aim to broaden the domain of the merged images by applying an ample range of intensity transformations $\mathcal{T}_{\text{intensity}}$ with loose configurations. We consecutively perform random bias field augmentations, add Gaussian noise, apply random localized spikes in k-space, randomly adjust the image contrast, perform Gaussian smoothing with either individual or shared $\sigma$ values for all spatial dimensions, add Rician noise, apply Gibbs noise, perform random Gaussian sharpening, and randomly transform intensity histograms.

\begin{table*}[t]
\centering
\scriptsize
\caption{
Quantitative results. We compare vesselFM to state-of-the-art foundation models for medical image segmentation on three tasks: \textit{zero}-, \textit{one}-, and \textit{few}-shot 3D blood vessel segmentation. VesselFM is evaluated on four datasets of clinical (SMILE-UHURA~\cite{chatterjee2024smile}, MSD8~\cite{antonelli2022medical}) and pre-clinical (OCTA~\cite{wittmann2024simulation,glandorf2024bessel}, BvEM~\cite{wan2024trisam}) relevance and consistently outperforms all baselines by a relatively large margin.
}
\label{tab:quantitative_results}
\begin{tabular}{c|l| c c| c c| c c| c c} 

\toprule
\multirow{2}{*}{Task} & \multirow{2}{*}{Model} &
\multicolumn{2}{c|}{OCTA~\cite{wittmann2024simulation,glandorf2024bessel}} & \multicolumn{2}{c|}{BvEM~\cite{wan2024trisam}} & \multicolumn{2}{c|}{SMILE-UHURA~\cite{chatterjee2024smile}}& \multicolumn{2}{c}{MSD8~\cite{antonelli2022medical}}\\
&& $\text{Dice}\uparrow$ & $\text{clDice}\uparrow$ & $\text{Dice}\uparrow$ & $\text{clDice}\uparrow$ & $\text{Dice}\uparrow$ & $\text{clDice}\uparrow$ & $\text{Dice}\uparrow$ & $\text{clDice}\uparrow$\\

\midrule
\multirow{5}{*}{\rotatebox[origin=c]{90}{\textit{zero-shot}}}   
& tUbeNet~\cite{holroyd2023tube} & 36.01 & 23.64 & 10.03 & 11.17 & 48.32 & 36.85 & 5.13 & 5.84\\
& VISTA3D~\cite{he2024vista3d} & 13.60 & 3.72 & 0.94 & 2.03 & 5.05 & 1.62 & 23.83 & 20.25\\
& SAM-Med3D~\cite{wang2024sammed3d} & 6.74 & 6.56 & 5.98 &  7.38 & 2.12 & 1.66 & 7.94 & 7.89\\
& MedSAM-2~\cite{zhu2024medical} & 28.56 & 15.76 & 10.92 & 12.27 & 3.85 & 5.46 & 14.53 & 14.27\\
\cmidrule{2-10}
& vesselFM (ours) & \textbf{46.94} & \textbf{67.07} & \textbf{67.49} & \textbf{62.04} & \textbf{74.66} & \textbf{75.27} & \textbf{29.69} & \textbf{36.14}\\

\midrule
\multirow{6}{*}{\rotatebox[origin=c]{90}{\textit{one-shot}}}   
& tUbeNet~\cite{holroyd2023tube} & 38.09 & 59.37 & 10.75 & 11.53 & 57.67 & 53.25 & 13.66 & 15.41\\
& VISTA3D~\cite{he2024vista3d} & 51.24 & 25.69 & 8.25 & 15.04 & 56.53 & 42.42 & 31.73 & 32.94\\
& SAM-Med3D~\cite{wang2024sammed3d} & 38.33 & 54.90 & 49.47 & 52.14 & 38.57  & 36.94 & 29.29 & 36.78\\
& MedSAM-2~\cite{zhu2024medical} & 56.68 & 50.95 &  24.07 & 24.69 & 19.78 & 11.87 & 30.21 & 23.89\\
\cmidrule{2-10}
& vesselFM (from scratch)$^{*}$ & 65.57 & 73.79 & 63.85 & 39.55 & 37.99 & 45.72 & 27.13 & 29.48\\
& vesselFM (ours) & \textbf{72.10} & \textbf{83.73} & \textbf{78.27} & \textbf{79.91} & \textbf{76.43} & \textbf{78.36} & \textbf{36.88} & \textbf{48.65}\\

\midrule
\multirow{6}{*}{\rotatebox[origin=c]{90}{\textit{few-shot}}}   
& tUbeNet~\cite{holroyd2023tube} & 41.61 & 57.98 &  5.41 & 10.22 & 56.31 & 49.28 & 17.67 & 18.97\\
& VISTA3D~\cite{he2024vista3d} & 54.25 & 32.59 & 24.04 & 38.10 & 61.17& 51.05 & 41.90 & 46.45\\
& SAM-Med3D~\cite{wang2024sammed3d} & 37.85 & 56.94 & 57.86 & 66.04 & 46.59 & 44.63 & 31.30 & 35.48\\
& MedSAM-2~\cite{zhu2024medical} & 56.96 & 51.99 & 18.76 & 19.66 & 58.15 & 42.72 & 29.24 & 22.38\\
\cmidrule{2-10}
& vesselFM (from scratch)$^{*}$ & 67.37 & 75.79 &63.03 & 56.69 & 50.51 & 58.77 & 32.64 & 36.03\\
& vesselFM (ours) & \textbf{75.70} & \textbf{84.03} & \textbf{78.11} & \textbf{84.54} & \textbf{78.77} & \textbf{79.37} & \textbf{45.04} & \textbf{57.25}\\
\bottomrule
\multicolumn{7}{l}{\rule{0pt}{2.5mm} \scriptsize{$^{*}$Model not pre-trained on $\mathcal{D}_\text{real}$, $\mathcal{D}_\text{drand}$, and $\mathcal{D}_\text{flow}$.}}
\end{tabular}
\end{table*}

\subsection{\texorpdfstring{$\mathcal{D}_\text{flow}$:}{} Flow Matching-Based Image Generation}
Flow matching~\cite{lipman2023flow,liu2022flow} is a promising alternative to diffusion models and has shown superior performance on natural images~\cite{esser2024scaling,ma2024sit}. To generate our third data source $\mathcal{D}_\text{flow}$ = $\{\mathcal{X}_\text{flow}, \mathcal{M}_\text{syn}\}$, we train and subsequently sample images from a mask- and class-conditioned flow matching-based generative model $\mathcal{F}$, aiming at further broadening the distributions of $\mathcal{D}_\text{real}$ in a data-driven manner. $\mathcal{F}$ utilizes a $\theta$-parametrized network representing a learned, time-dependent velocity field $v$, which is trained to map samples $x_{0} \sim \mathcal{N}(0, I)$ to samples $x_{1}$ of the data distribution via an ordinary differential equation (ODE):
\begin{equation}\label{eq:ode}
\frac{\mathrm{d}}{\mathrm{d}t}x_t = v_{\theta}(x_t,m,c,t),
\end{equation}
where $t \in [0, 1]$ represents the time, $c$ the class we condition on, and $m$ the conditioning mask. To train $\mathcal{F}$, we optimize the conditional flow matching (CFM) objective~\cite{lipman2023flow}, which minimizes the $L_2$ loss between the predicted velocity $v_{\theta}(x_t,m,c,t)$ and the sampled ground truth velocity $u_t(x_t|x_1)$ at time $t$:
\begin{equation}
\mathcal{L}_\text{CFM}(\theta) = \mathbb{E}_{t, x_1, x_t} \| v_{\theta}(x_t,m,c,t) - u_t(x_t|x_1) \|^2.
\end{equation}
We define the forward process as $x_t=t x_1 + (1-t) x_0$, leading to $u_t(x_t|x_1) = (x_1 - x_t) / (1-t)$ in the loss above. The time-linear forward process provides straighter ODE trajectories than a popular variance-preserving diffusion noise schedule in DDPM, simplifying sampling at inference~\cite{liu2022flow,lipman2023flow,ma2024sit}. We train $\mathcal{F}$ on matching image-mask pairs $(x_1, m)$ and their associated classes $c \in \mathcal{C} \cup \{ 0 \}$, sampled from \textit{both} of our previously generated data sources.
Building on anatomically controllable medical image generation methods~\cite{dorjsembe2024conditional,konz2024anatomically}, we implement mask conditioning by concatenating the mask channel-wise with the input image $x_t$. Class information is incorporated by adding the class embedding to the time embedding, followed by injection into the intermediate feature layers via addition.

\noindent
To generate $\mathcal{D}_\text{flow}$, we ultimately sample a vast amount of images $\mathcal{X}_\text{flow}$ (see Fig.~\ref{fig:method_fm}) by discretizing (\ref{eq:ode}) via Euler integration:
\begin{equation}\label{eq:sampling}
x_{t+\Delta t} = x_t + v_{\theta}(x_t,m,c,t) \cdot \Delta t, \quad \Delta t = \frac{1}{N},
\end{equation}
where $N$ represents the total number of time steps. Given that $\mathcal{M}_\text{syn}$ covers the required range of blood vessel patterns and is devoid of annotator-induced biases and errors in segmentation masks, we opt to exclusively use masks $m \sim \mathcal{M}_\text{syn}$ during sampling. For clarity, we use tilde to denote classes of data generated by $\mathcal{F}$ (\eg, $\Tilde{7}$).

\section{Experiments and Results}\label{sec:experiments}
In this section, we elaborate on our findings and showcase vesselFM's performance on three tasks: \textit{zero}-shot, \textit{one}-shot, and \textit{few}-shot segmentation. We evaluate vesselFM on four 3D blood vessel segmentation datasets of unseen clinically (SMILE-UHURA~\cite{chatterjee2024smile}, MSD8~\cite{antonelli2022medical}) and pre-clinically (OCTA~\cite{wittmann2024simulation,glandorf2024bessel}, BvEM~\cite{wan2024trisam}) relevant domains. In this context, we extract three patches of shape $\text{128}^\text{3}$ from each of these evaluation datasets and use the remaining data for testing and validation (see Suppl.~\ref{suppl:preprocessing} for details). With the three extracted patches, we define the \textit{one}- and \textit{few}-shot segmentation task as fine-tuning models on either one or all three patches. For \textit{zero}-shot evaluation, we apply models out-of-the-box on the test data without prior fine-tuning. This setup mimics clinical scenarios where annotated data is scarce.

We compare vesselFM with four foundation models designed for 3D medical image segmentation: the generalizable 3D blood vessel segmentation model tUbeNet~\cite{holroyd2023tube}, the CT-specific VISTA3D~\cite{he2024vista3d}, and the two SAM-like general-purpose segmentation models SAM-Med3D~\cite{wang2024sammed3d} and MedSAM-2~\cite{zhu2024medical}. In our experiments, we exclude the classes of the four datasets used for evaluation (see Table~\ref{tab:datasets}, upper section) from the curation of $\mathcal{D}_\text{real}$ and $\mathcal{D}_\text{flow}$. 
We generate $\mathcal{D}_\text{flow}$ by sampling 10,000 image-mask pairs from $\mathcal{F}$ on a single RTX A6000 GPU over the course of three days. To curate $\mathcal{D}_\text{drand}$, we sample 500,000 image-mask pairs from our domain randomized generative pipeline. All images-mask pairs are of shape $\text{128}^{\text{3}}$. We train vesselFM using all three data sources, with weights assigned roughly according to their sizes (70\% $\mathcal{D}_\text{drand}$, 20\% $\mathcal{D}_\text{real}$, and 10\% $\mathcal{D}_\text{flow}$). We opt for MONAI's reimplementation of the UNet architecture proposed by Isensee \etal~\cite{isensee2021nnu} to present our segmentation model. For flow matching, we use the UNet from Med-DDPM~\cite{dorjsembe2024conditional} to represent the learned velocity field $v$. We set the total number of time steps $N$ in (\ref{eq:sampling}) to 100. Following common practices, we report Dice scores and topology-aware centerline Dice (clDice) scores~\cite{shit2021cldice}, which assess the preservation of tubular appearance and connectivity of blood vessels.
Further details on our experimental setup can be found in Suppl.~\ref{suppl:exp_setup}.

\begin{figure*}[ht]
\centerline{\includegraphics[width=\linewidth]{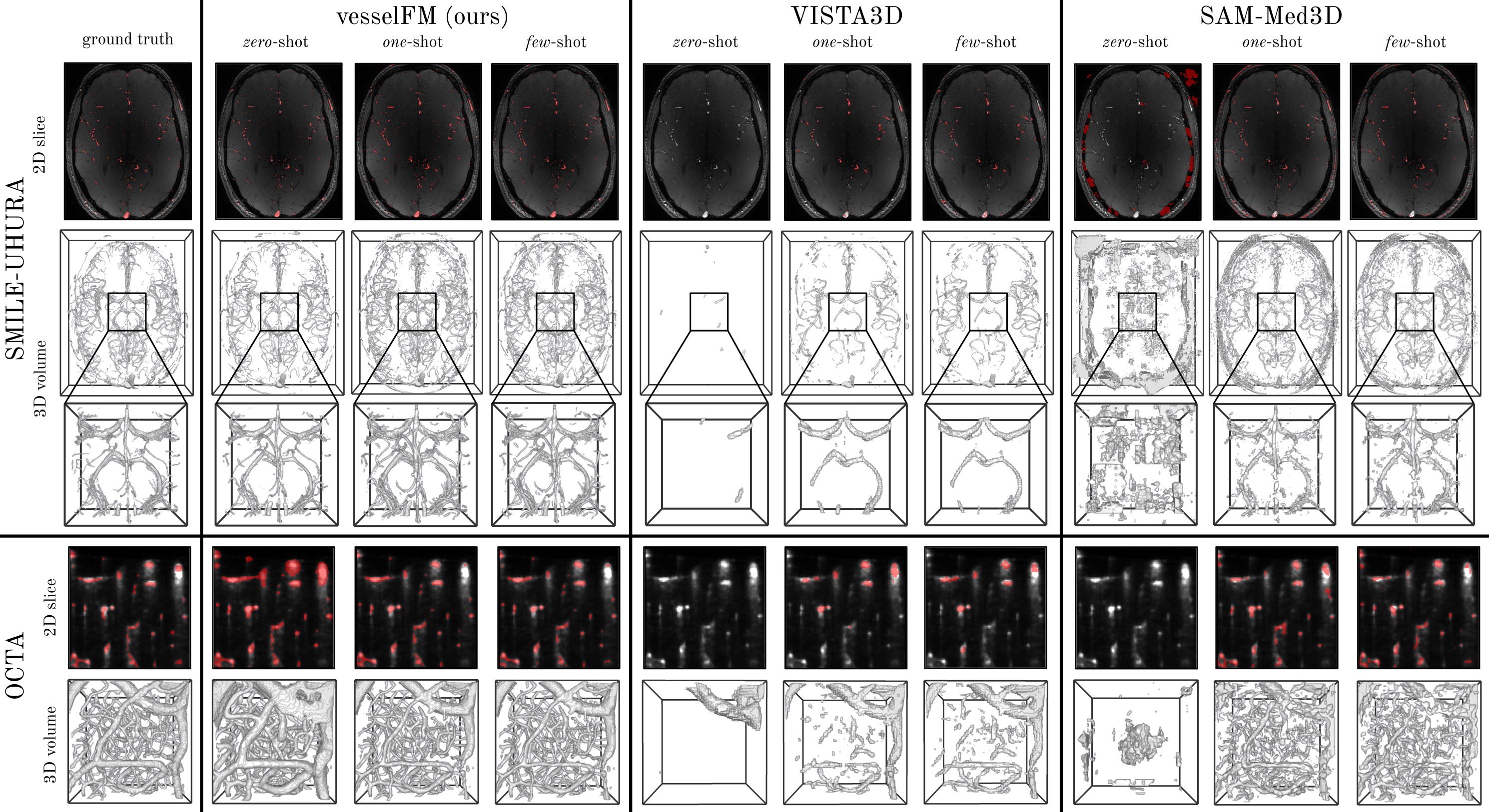}}
\caption{
Qualitative results (better viewed zoomed in). We visualize predictions on the SMILE-UHURA and the OCTA datasets for all three tasks (\textit{zero}-, \textit{one}-, and \textit{few}-shot). We compare vesselFM to VISTA3D, SAM-Med3D, and ground truth segmentation masks. Note that vesselFM consistently predicts state-of-the-art results, even in the \textit{zero}-shot setting, demonstrating exceptional generalization to unseen domains. For better visibility, we show a zoomed-in view of the 3D predictions on the SMILE-UHURA datasets.
}
\label{fig:qual_res}
\end{figure*}

\subsection{Quantitative and Qualitative Results}\label{sec:quant_qual_res}
Quantitative results can be observed in Table~\ref{tab:quantitative_results}. We find that our proposed foundation model, vesselFM, tailored to universal 3D blood vessel segmentation, outperforms the baseline models on all datasets and tasks by a large margin.

\paragraph{\textit{Zero}-shot task.}
VesselFM exhibits exceptional \textit{zero}-shot generalization on all four datasets, which cover a diverse array of unseen domains and even imaging modalities (OCTA and vEM). Surprisingly, vesselFM scores 5.86 Dice points higher than VISTA3D on MSD8, even though VISTA3D was trained on 11,454 CT volumes, including data from MSD8 itself. This highlights vesselFM's strong inductive bias, enabled by training on our three proposed heterogeneous blood vessel data sources. Moreover, vesselFM outperforms the generalizable 3D blood vessel segmentation model tUbeNet, trained on four blood vessel datasets of varying imaging modalities. We observe that tUbeNet struggles in more complex imaging modalities where blood vessels do not have a stark contrast to background tissues (\eg, BvEM and MSD8). The general-purpose segmentation models SAM-Med3D and MedSAM-2 both fail to segment blood vessels in the \textit{zero}-shot setting. Notably, vesselFM's \textit{zero}-shot results surpass \textit{few}-shot results achieved by baseline models on SMILE-UHURA in Dice and clDice.

\paragraph{\textit{One}- and \textit{few}-shot tasks.}
Fine-tuning vesselFM in a \textit{one}- or \textit{few}-shot manner generally increases segmentation performance. Given that some baseline models rely on networks, which may easily overfit to the small amount of training data provided in the \textit{one}- and \textit{few}-shot settings, we additionally compare vesselFM to a variant of the same configuration that is trained from scratch without being pre-trained on our three proposed data sources (see footnote in Table~\ref{tab:quantitative_results}; see Suppl.~\ref{suppl:scratch} for more insights). We observe that neglecting our three proposed data sources causes a notable decrease in Dice and clDice scores, validating our rationale.

Qualitative results mirror the insights gained from quantitative results. Specifically, we find that vesselFM demonstrates exceptional \textit{zero}-shot generalizations, free of annotator-specific biases (see Fig.~\ref{fig:qual_res}).

\subsection{Ablation Studies}
We ablate vesselFM's design choices on the SMILE-UHURA~\cite{chatterjee2024smile} dataset, targeting human brain vessel segmentation in MRAs, a task of high clinical importance for automated diagnosis of various diseases, such as aneurysms. All ablations are conducted for the \textit{zero}-shot segmentation task, given that \textit{zero}-shot generalization is the most pivotal component of segmentation foundation models.

\begin{figure}[ht]
\centerline{\includegraphics[width=\linewidth]{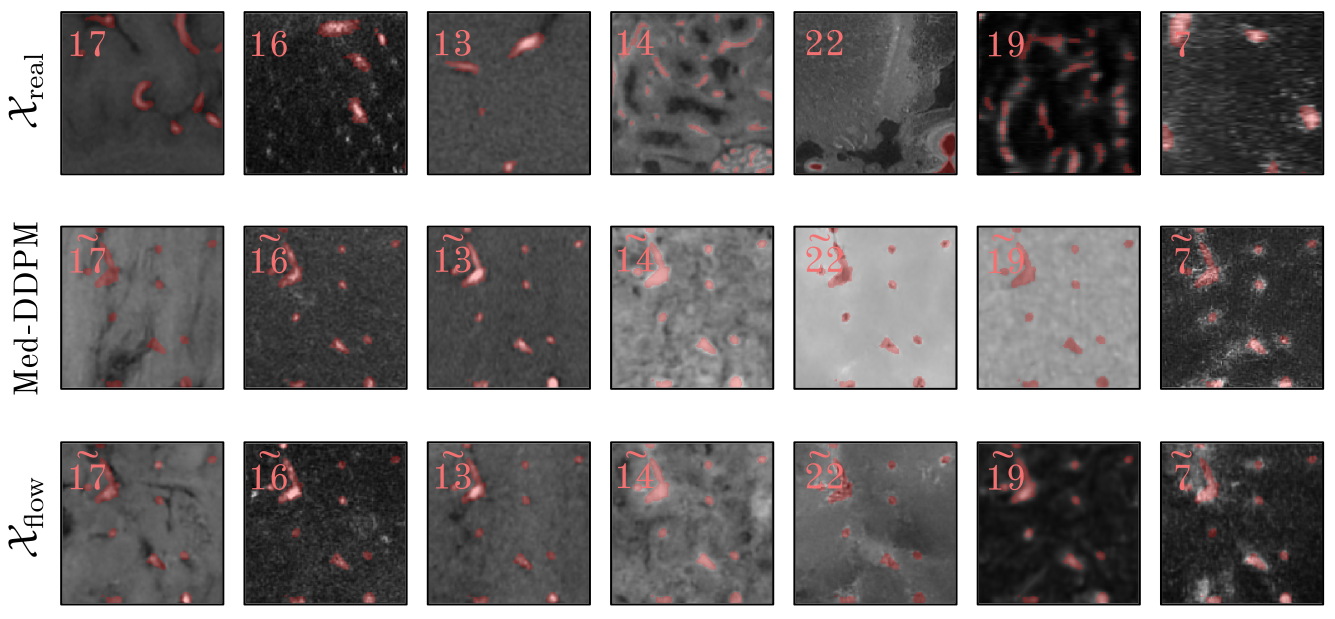}}
\caption{
Qualitative comparison of images generated by our flow matching-based generative model $\mathcal{F}$ ($\mathcal{X}_\text{flow}$, 3rd row) with images generated by the diffusion-based generative model Med-DDPM~\cite{dorjsembe2024conditional} (2nd row). We also include real images of the same classes for reference ($\mathcal{X}_\text{real}$, 1st row). For improved comparability, we consistently condition $\mathcal{F}$ and Med-DDPM on the same mask $m$. Segmentation masks are displayed in translucent red.
}
\label{fig:ddpm_vs_fm}
\end{figure}

First, we ablate the relevance of our three proposed data sources, $\mathcal{D}_\text{real}$, $\mathcal{D}_\text{drand}$, and $\mathcal{D}_\text{flow}$ (see Table~\ref{tab:ablations_data_sources}). To this end, we train vesselFM on each data source individually and progressively augment $\mathcal{D}_\text{real}$ with $\mathcal{D}_\text{drand}$ and $\mathcal{D}_\text{flow}$. A more complete ablation of our data sources covering all four evaluation datasets and all tasks can be found in Suppl.~\ref{suppl:source_abl}.
\begin{table}[ht]
\centering
\scriptsize
\caption{Ablation of data sources.}
\label{tab:ablations_data_sources}
\begin{tabular}{l|c c} 
\toprule
Data sources & $\text{Dice}\uparrow$ & $\text{clDice}\uparrow$\\
\midrule
$\mathcal{D}_\text{real}$ & 65.45 & 63.53\\
$\mathcal{D}_\text{real}$ + $\mathcal{D}_\text{drand}$ & 69.38 & 72.10\\
$\mathcal{D}_\text{real}$ + $\mathcal{D}_\text{drand}$ + $\mathcal{D}_\text{flow}$& \textbf{74.66} & \textbf{75.27}\\
\midrule
$\mathcal{D}_\text{drand}$ & 55.34 & 63.16\\
$\mathcal{D}_\text{flow}$ & 14.33 & 16.14\\
\bottomrule
\end{tabular}
\end{table}
We observe that supplementing $\mathcal{D}_\text{real}$ with $\mathcal{D}_\text{drand}$ and $\mathcal{D}_\text{flow}$ results in an impressive increase in Dice and clDice scores of 9.21 and 11.74, respectively. This highlights that leveraging all our three proposed heterogeneous data sources collaboratively allows vesselFM to learn robust features that are well-suited for \textit{zero}-shot generalization, enabling a foundation model for universal 3D blood vessel segmentation.

Second, we ablate the design choices of our proposed mask- and class-conditioned flow matching-based generative model $\mathcal{F}$, used to sample $\mathcal{D}_\text{flow}$ (see Table~\ref{tab:ablations_flow}).
Specifically, we exclude $\mathcal{D}_\text{drand}$ from training of $\mathcal{F}$ (2nd row), utilize real masks $\mathcal{M}_\text{real}$ instead of our proposed synthetic masks $\mathcal{M}_\text{syn}$ during sampling from $\mathcal{F}$ (3rd row), and exclude class conditioning (4th row). Subsequently, we replace $\mathcal{D}_\text{flow}$ with the respective generated variants and re-train vesselFM.
\begin{table}[ht]
\centering
\scriptsize
\caption{Ablation of mask- and class-conditioned flow matching.}
\label{tab:ablations_flow}
\begin{tabular}{l|c c} 
\toprule
Method used for $\mathcal{D}_\text{flow}$ & $\text{Dice}\uparrow$ & $\text{clDice}\uparrow$\\
\midrule
$\mathcal{F}$ & \textbf{74.66} & \textbf{75.27}\\
$\mathcal{F}$, no $\mathcal{D}_\text{drand}$ & 71.24 & 73.93\\
$\mathcal{F}$, $\mathcal{M}_\text{real}$ & 70.12 & 70.84\\
$\mathcal{F}$, no class cond. & 74.56 & 74.75\\
\midrule    
Med-DDPM~\cite{dorjsembe2024conditional} & 70.34 & 73.35	\\
\bottomrule
\end{tabular}
\end{table}
Omitting $\mathcal{D}_\text{drand}$, and consequently the class $c = 0$, from training reduces the Dice score by 3.42. This suggests that the added data diversity and quantity provided by $\mathcal{D}_\text{drand}$ enables $\mathcal{F}$ to generate a wider variety of images, thereby facilitating generalization. Generating $\mathcal{D}_\text{flow}$ by conditioning on real masks $\mathcal{M}_\text{real}$ included in $\mathcal{D}_\text{real}$ instead of synthetic masks results in a 4.54 reduction in Dice score. This finding supports our decision to utilize synthetic masks sampled from $\mathcal{M}_\text{syn}$, which are, unlike masks from  $\mathcal{M}_\text{real}$, free of annotator-induced errors and offer greater diversity. We find that omitting class conditioning leads to a modest drop in Dice by 0.10.
Finally, we compare $\mathcal{F}$, relying on the concept of flow matching, to the diffusion-based, generative model Med-DDPM~\cite{dorjsembe2024conditional} (5th row). $\mathcal{F}$ outperforms Med-DDPM not only quantitatively (4.32 Dice) but also qualitatively (see Fig.~\ref{fig:ddpm_vs_fm}). Specifically, we observe that Med-DDPM frequently struggles to capture class-specific artifacts accurately, resulting in low-fidelity synthetic images (\eg, see Fig.~\ref{fig:ddpm_vs_fm}, classes 22 and 19).

Third, we ablate vesselFM's segmentation model by experimenting with relevant medical 3D segmentation networks (see Table~\ref{tab:ablations_arch}).
\begin{table}[ht]
\centering
\scriptsize
\caption{Ablation of vesselFM's segmentation model.}
\label{tab:ablations_arch}
\begin{tabular}{l |c c c} 
\toprule
Segmentation arch. & $\text{Dice}\uparrow$ & $\text{clDice}\uparrow$\\
\midrule
UNet &  \textbf{74.66} & \textbf{75.27}\\
SwinUNETR~\cite{hatamizadeh2021swin} &  60.00 & 53.92 \\
SwinUNETR-V2~\cite{he2023swinunetr} &  74.54 & 74.80\\
UNETR~\cite{hatamizadeh2022unetr} & 46.74 & 40.15\\
3D UX-Net~\cite{lee2023d} & 49.99 & 46.31\\
MedNeXt~\cite{roy2023mednext} & 56.47 & 61.95\\
\bottomrule
\end{tabular}
\end{table}
Our employed UNet variant surpasses transformer-based~\cite{hatamizadeh2021swin,he2023swinunetr,hatamizadeh2022unetr} and ConvNeXt~\cite{liu2022convnet}-based~\cite{lee2023d,roy2023mednext} networks alike, which accurately represents the current landscape in medical image segmentation.
\section{Conclusion and Outlook}\label{sec:conclusion}
In this work, we propose vesselFM, a foundational model for universal 3D blood vessel segmentation. VesselFM is capable of accurately segmenting 3D vasculature in previously unseen modalities and tissue types and performs superior to state-of-the-art medical image segmentation foundational models.
We enable \textit{zero}-shot generalization by training vesselFM on three proposed heterogeneous data sources ($\mathcal{D}_\text{real}$, $\mathcal{D}_\text{drand}$, and $\mathcal{D}_\text{flow}$), which we extensively ablate in our experiments. Given that vesselFM pushes the frontier in the (pre-)clinically relevant task of 3D blood vessel segmentation, we hope that our work enables novel insights into vascular disorders and fosters the development of advanced diagnostic tools, ultimately resulting in improved patient outcomes (see Suppl.~\ref{suppl:clinical} for discussion on clinical utility). We advise future work to experiment with tailored post-processing steps to improve blood vessel connectivity (potentially at the graph level~\cite{wittmann2024link}) and extend vesselFM to multi-class or instance segmentation tasks.

\paragraph{Acknowledgments:} This work was supported by the Helmut Horten Foundation. \textit{TubeTK dataset}: The MR brain images from healthy volunteers used in this paper were collected and made available by the CASILab at The University of North Carolina at Chapel Hill and were distributed by the MIDAS Data Server at Kitware, Inc.

{
    \small
    \bibliographystyle{ieeenat_fullname}
    \bibliography{main}
}

\appendix
\clearpage
\setcounter{page}{1}
\setcounter{table}{5}
\setcounter{figure}{7}
\maketitlesupplementary

\begin{figure}[h]
\centerline{\includegraphics[width=\linewidth]{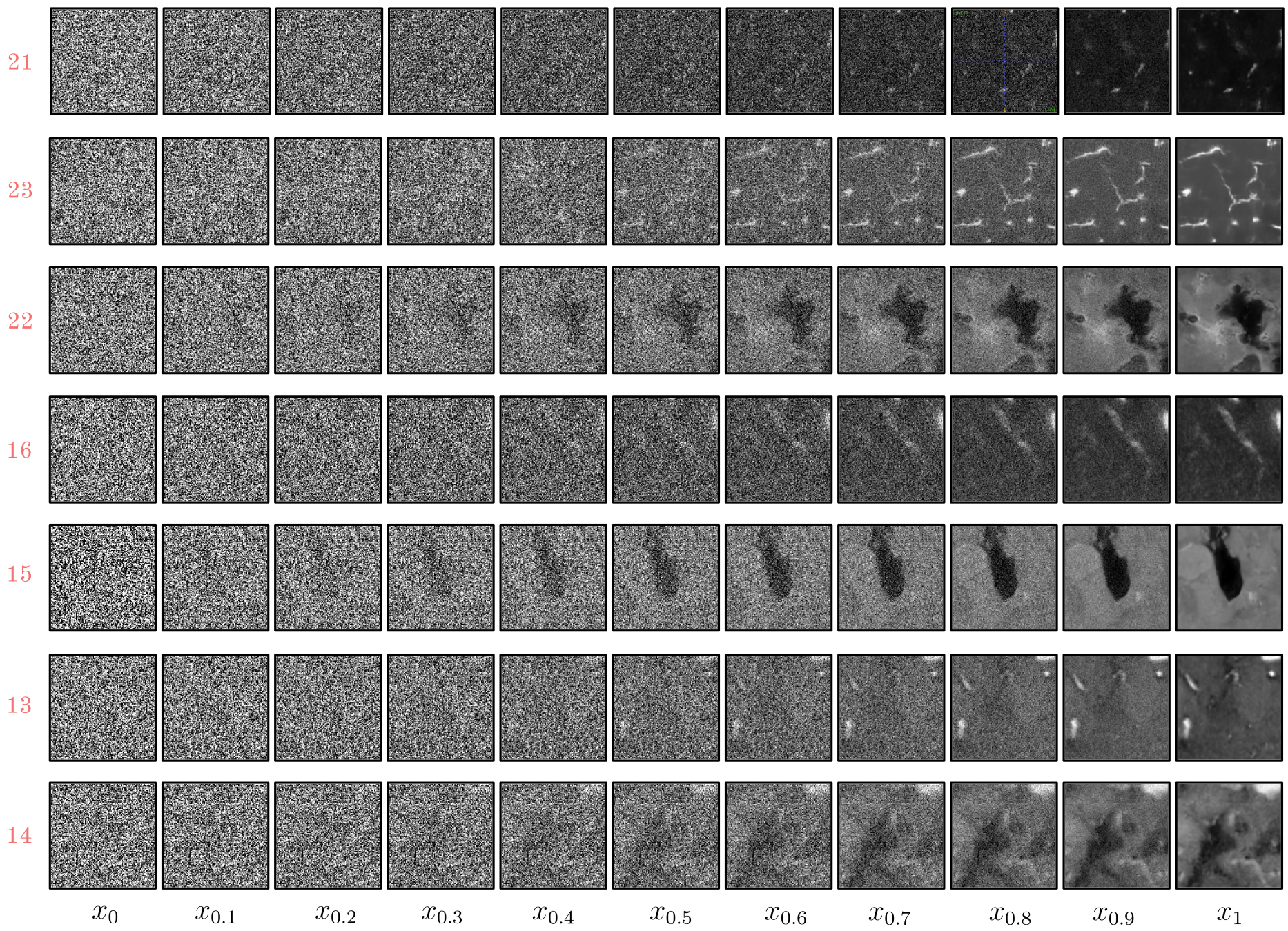}}
\caption{
More trajectories of flow matching-based sampling. For improved visibility, we plot 2D slices. Given that the total number of time steps $N$ is set to 100 in our experiments, $\Delta t$ of 0.1 corresponds to 10 steps. Classes are indicated in red.
}
\label{fig:suppl_trajectories}
\end{figure}

\begin{figure}[t]
\centerline{\includegraphics[width=\linewidth]{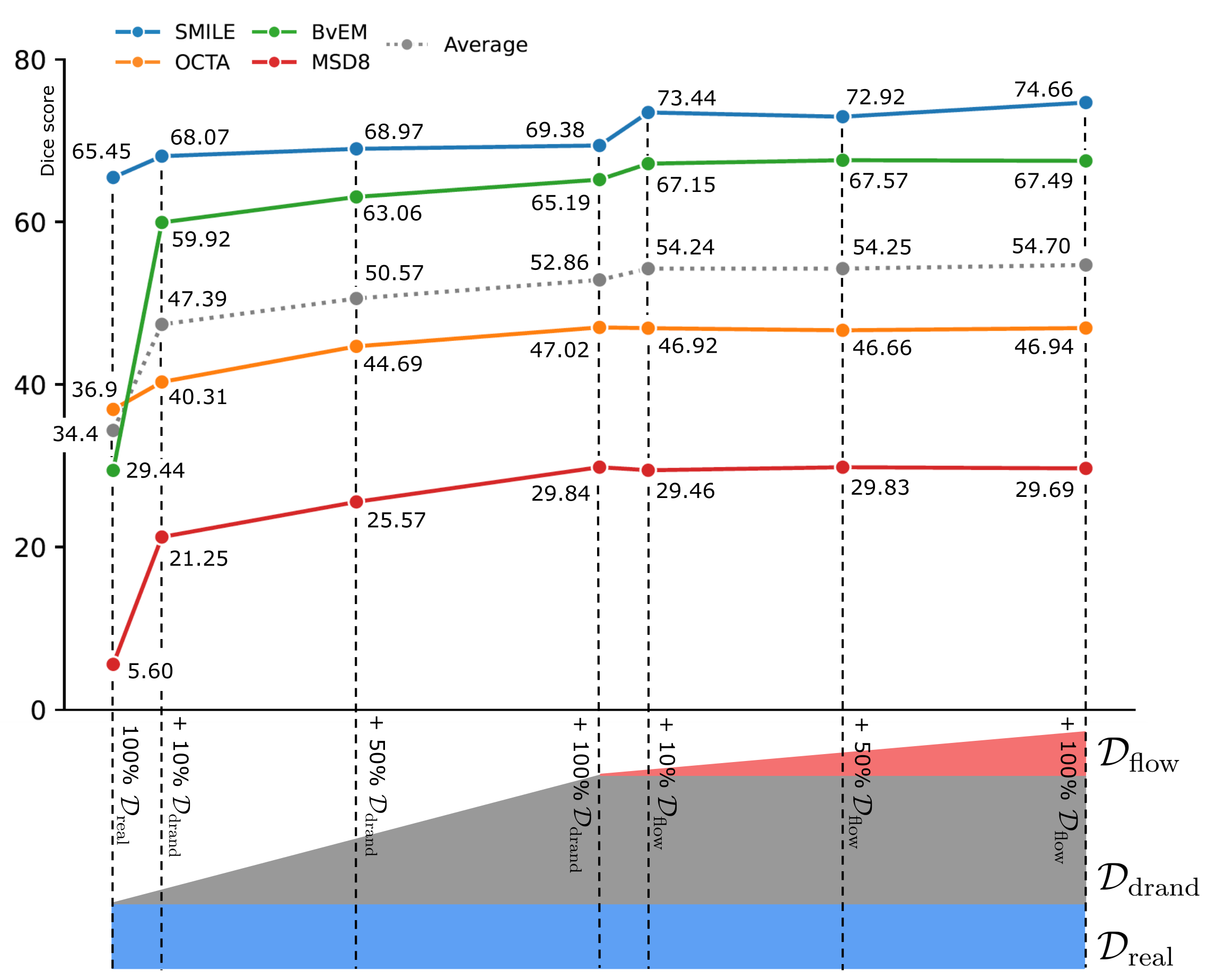}}
\caption{ 
Experiment on the effect of our three proposed data sources on vesselFM's performance. We gradually augment $\mathcal{D}_{\text{real}}$ with 10\%, 50\%, and 100\% of $\mathcal{D}_{\text{drand}}$, followed by adding 10\%, 50\%, and 100\% of $\mathcal{D}_{\text{flow}}$ (see bottom part). We report \textit{zero}-shot Dice scores on the four evaluation datasets. We generally find that augmenting $\mathcal{D}_{\text{real}}$ with $\mathcal{D}_{\text{drand}}$ and $\mathcal{D}_{\text{flow}}$ results in increased segmentation performance (see average).
}
\label{fig:suppl_amount_of_training_data}
\end{figure}

\begin{figure*}[ht]
\centering
\includegraphics[width=0.9\linewidth]{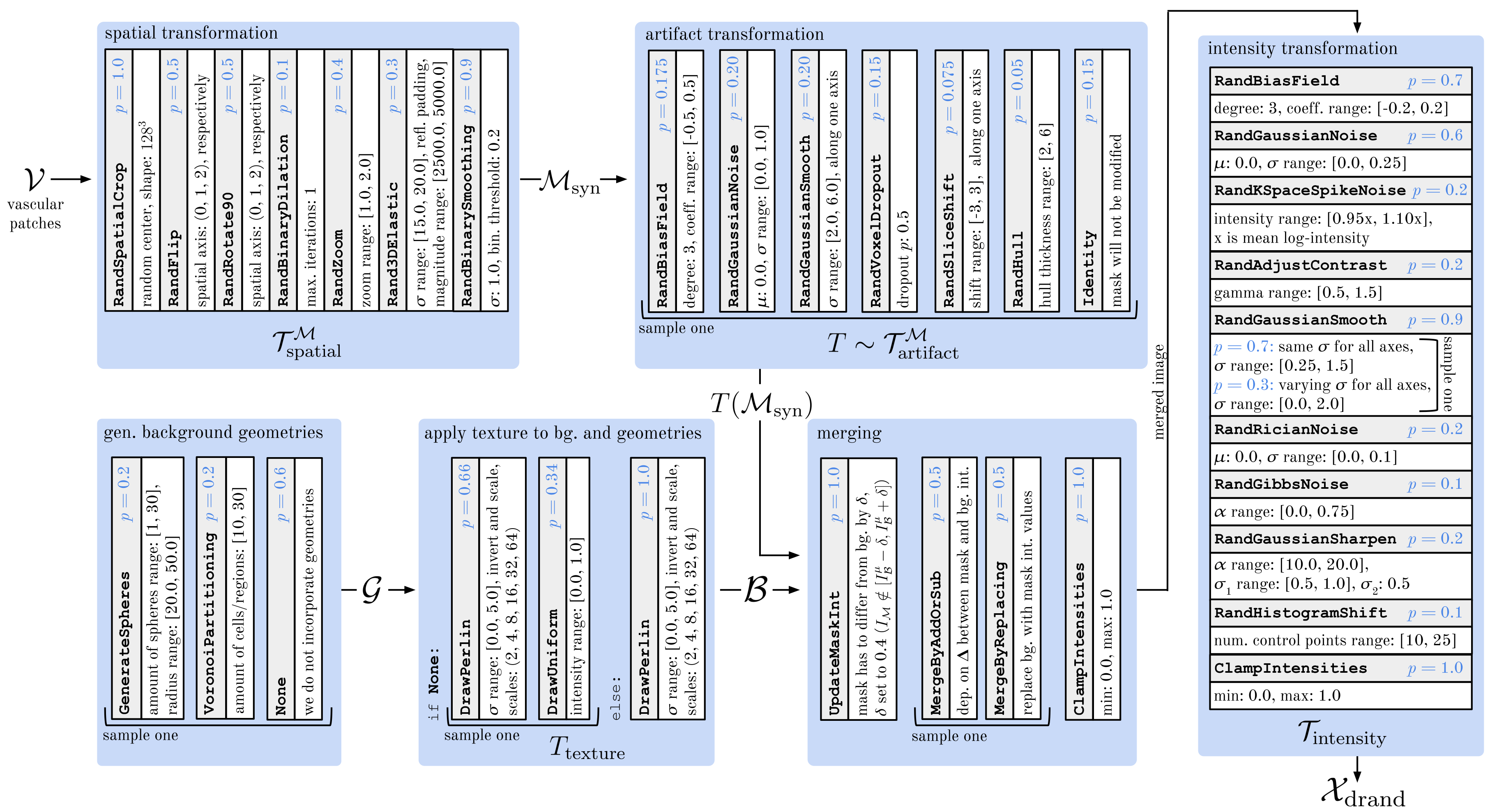}
\caption{
Parametrization of our domain randomized generative pipeline. All parameters were carefully tuned to ensure sufficient diversity while preserving key characteristics relevant to the general domain of vascular images. If not indicated otherwise, all the above transformations are applied consecutively, starting from the left-hand side. Probabilities associated with specific transformations are indicated in blue. Following common practices in medical image analysis, we utilize, whenever possible, transformations from the MONAI framework. Note that we exactly follow the notation from Fig.~\ref{fig:method_drand}.
}
\label{fig:suppl_drand_param}
\end{figure*}

\section{Pre-Processing of Datasets in \texorpdfstring{$\mathcal{D}_\text{real}$}{}}\label{suppl:preprocessing}
\paragraph{General pre-processing steps.}
We aim to ensure that all 23 datasets used to curate $\mathcal{D}_\text{real}$ (see Table~\ref{tab:datasets}) adhere to general vascular imaging characteristics and comply with our label quality standards. To this end, we apply carefully selected pre-processing steps (see Table~\ref{tab:datasets}, last column).
First, we resample the BvEM~\cite{wan2024trisam}, TubeTK~\cite{bullitt2005vessel}, tUbeNet~\cite{holroyd2023tube}, TopCoW~\cite{yang2023benchmarking}, and DeepVesselNet~\cite{tetteh2020deepvesselnet} datasets, establishing appropriate blood vessel scales (\eg, a single vessel should not occupy 90\% of the patch) and near-isotropic voxel sizes. In the case of the BvEM dataset, only the labels are resampled, as the annotations were likely made on a downsampled version of the volume.
As resampling may introduce label artifacts, we subsequently smooth affected labels (TubeTK and tUbeNet) using Gaussian smoothing followed by thresholding.
Since the 3D-IRCADb-01~\cite{soler20103d} dataset contains labels of multiple structures, we solely keep venous system, artery, and portal vein labels, converting them to binary labels. The original labels of the HR-Kindney~\cite{kuo2023terabyte} dataset are of relatively poor quality. However, enabled by the high signal-to-noise ratio of the volume representing the HR-Kindney dataset, we improve label quality by applying Algorithm~\ref{alg:hrlabels}.
To further enhance the visibility of blood vessels, the intensities of the VesSAP~\cite{todorov2020machine} and LS~\cite{binder2024leptomeningeal} datasets are clipped at the 0\% and 98\% percentiles, while the intensities of the MSD8 dataset are clipped at the 20\% and 98\% percentiles.
We crop the MSD8 and 3D-IRCADb-01 datasets to retain only foreground structures, as the images otherwise would primarily consist of non-annotated anatomical structures. Additionally, we crop the borders from the BvEM volume, given that they predominantly contain artifacts.

\begin{algorithm}
\caption{HR-Kidney label improvement.}\label{alg:hrlabels}
\scriptsize{
\textbf{Input:} Image, Intensity Delta = 0.1, Threshold = 0.9, Filter Size = 11
\begin{algorithmic}
\State $\text{Median} \gets \text{MedianFilter(Image, Filter Size)}$ \Comment{Apply filter to image.}
\State $\text{Mask} \gets \text{(Image - Median)} > \text{Int. Delta}$ \Comment{Include high local int. variations.}
\State $\text{Mask} \gets \text{Mask} \lor (\text{Image} > \text{Threshold}$) \Comment{Include high int. values.}
\State $\text{Mask} \gets \text{Mask} \bullet \mathbf{1}_{3\times3\times3}$   \Comment{Close small gaps.}
\State $\text{Mask} \gets \text{RemoveSmallObjects(Mask)}$ \Comment{Remove small connected components.}
\end{algorithmic}
}
\scriptsize{$\bullet$ denotes morphological closing.}
\end{algorithm}

\paragraph{Evaluation datasets.}
From each evaluation dataset (see Table~\ref{tab:datasets}, upper section), we extract three patches of size $\text{128}^\text{3}$ for fine-tuning models in the \textit{one}- and \textit{few}-shot settings and use the remaining data for testing and validation.
\noindent
\textit{OCTA}~\cite{wittmann2024simulation,glandorf2024bessel}:
We allocate three of the six samples provided in the OCTA dataset for \textit{one}- and \textit{few}-shot fine-tuning and reserve the remaining three for model evaluation: two samples for testing and one sample for validation. The three samples used for fine-tuning are center-cropped to adhere to our target shape of $\text{128}^{\text{3}}$.
\noindent
\textit{SMILE-UHURA}~\cite{chatterjee2024smile}:
The fourteen samples in the SMILE-UHURA dataset are divided into one for validation, ten for testing, and three for extracting patches for fine-tuning. For the extraction of the fine-tuning patches, we pay special attention to extract patches highly representative of the characteristics of MRA scans contained in the SMILE-UHURA dataset (\eg, they contain vasculature, brain tissue, skull, and gyri/sulci).
\noindent
\textit{MSD8}~\cite{antonelli2022medical}:
We split the MSD8 dataset into one validation sample, 296 test samples, and three samples utilized to extract patches for fine-tuning. The patches for fine-tuning are chosen from representative regions and padded, if necessary, minimally in the $z$-dimension using reflective padding to conform to the target shape of $\text{128}^{\text{3}}$.
\noindent
\textit{BvEM}~\cite{wan2024trisam}:
The BvEM dataset contains solely a single volume of shape $\text{3571} \times \text{5145} \times \text{2495}$. We choose the first 130 slices to extract three $\text{128}^{\text{3}}$ patches for fine-tuning and one for validation. Then, we leave a buffer of 120 slices to minimize information leakage between the patches used for fine-tuning and testing. Lastly, we extract three bigger test volumes of shape $\text{500}^\text{3}$ from the remaining volume, limiting the overlap of these volumes with the fine-tuning patches in the $x$- and $y$-position as much as possible.

\section{Parameters of Domain Randomization}\label{suppl:param_drand}
An overview of the parametrization of transformations and operations in our proposed domain randomized generative pipeline, used to generate $\mathcal{D}_\text{drand}$, is shown in Fig.~\ref{fig:suppl_drand_param}.

\begin{table*}[t]
\centering
\scriptsize
\caption{
More detailed ablation of supplementing $\mathcal{D}_\text{real}$ with $\mathcal{D}_\text{drand}$ and $\mathcal{D}_\text{flow}$, covering all four evaluation datasets and all tasks. We generally find that the combination of all of our three proposed data sources yields the best segmentation performance.
}
\label{tab:suppl_ablation_data_source}
\begin{tabular}{c|l| c c| c c| c c| c c} 
\toprule
\multirow{2}{*}{\textit{Task}} & \multirow{2}{*}{Model} &
\multicolumn{2}{c|}{OCTA~\cite{wittmann2024simulation,glandorf2024bessel}} & \multicolumn{2}{c|}{BvEM~\cite{wan2024trisam}} & \multicolumn{2}{c|}{SMILE-UHURA~\cite{chatterjee2024smile}}& \multicolumn{2}{c}{MSD8~\cite{antonelli2022medical}}\\
&& $\text{Dice}\uparrow$ & $\text{clDice}\uparrow$ & $\text{Dice}\uparrow$ & $\text{clDice}\uparrow$ & $\text{Dice}\uparrow$ & $\text{clDice}\uparrow$ & $\text{Dice}\uparrow$ & $\text{clDice}\uparrow$\\
\midrule
\multirow{3}{*}{\rotatebox[origin=c]{90}{\tiny{\textit{zero-shot}}}}   
&$\mathcal{D}_\text{real}$ & 36.94 & 57.23 & 29.44 & 52.71 & 65.45 & 63.53 & 5.60 & 8.60\\
&$\mathcal{D}_\text{real}$ + $\mathcal{D}_\text{drand}$ & \textbf{47.02} & 61.05 & 65.19 & \textbf{65.13} & 69.38 & 72.10 & \textbf{29.84} & \textbf{37.47}\\
&$\mathcal{D}_\text{real}$ + $\mathcal{D}_\text{drand}$ + $\mathcal{D}_\text{flow}$ & 46.94 & \textbf{67.07} & \textbf{67.49} & 62.04 & \textbf{74.66} & \textbf{75.27} & 29.69 & 36.14\\
\midrule
\multirow{3}{*}{\rotatebox[origin=c]{90}{\tiny{\textit{one-shot}}}}   
&$\mathcal{D}_\text{real}$ & 69.32 & 77.68 & 72.01 & \textbf{85.22} & 72.20 & 74.87 & 27.14	& 40.51\\
&$\mathcal{D}_\text{real}$ + $\mathcal{D}_\text{drand}$ & 70.63	& 81.11 & 75.77	& 78.48	 & 71.77 & 71.90 & 35.35 & \textbf{49.39}\\
&$\mathcal{D}_\text{real}$ + $\mathcal{D}_\text{drand}$ + $\mathcal{D}_\text{flow}$ & \textbf{72.10} & \textbf{83.73} & \textbf{78.27} & 79.91 & \textbf{76.43} & \textbf{78.36} & \textbf{36.88} & 48.65\\
\midrule
\multirow{3}{*}{\rotatebox[origin=c]{90}{\tiny{\textit{few-shot}}}}   
&$\mathcal{D}_\text{real}$ & 73.01 & 80.14 & 67.18 & 81.41 & 77.63 & 77.24 & 38.65 & 48.71\\
&$\mathcal{D}_\text{real}$ + $\mathcal{D}_\text{drand}$ & 74.44 & 82.64 & 73.43 & \textbf{84.75} & 77.37 & 78.28 & 42.31 & 54.44\\
&$\mathcal{D}_\text{real}$ + $\mathcal{D}_\text{drand}$ + $\mathcal{D}_\text{flow}$ & \textbf{75.70} & \textbf{84.03} & \textbf{78.11} & 84.54 & \textbf{78.77} & \textbf{79.37} & \textbf{45.04} & \textbf{57.25}\\
\bottomrule
\end{tabular}
\end{table*}

\section{More Details on Experimental Setup}\label{suppl:exp_setup}
\paragraph{Parametrization of segmentation model.}
As already stated in the main manuscript, we opt for MONAI's re-implementation of \cite{isensee2021nnu}'s UNet architecture, called \texttt{DynUNet}, to present our segmentation model. To be specific, we set \texttt{strides} to [[1, 1, 1], [2, 2, 2], [2, 2, 2], [2, 2, 2], [2, 2, 2], [2, 2, 2]], \texttt{kernel\_size} to [[3, 3, 3], [3, 3, 3], [3, 3, 3], [3, 3, 3], [3, 3, 3], [3, 3, 3]], \texttt{upsample\_kernel\_size} to [[2, 2, 2], [2, 2, 2], [2, 2, 2], [2, 2, 2], [2, 2, 2]], \texttt{filters} to [32, 64, 128, 256, 320, 320], and activated residual connection-based convolution block (\texttt{res\_block}). We found that this parametrization performed best in our experiments.

\paragraph{Training of vesselFM.}
We employ a combination of Dice and cross-entropy loss functions, weighted by 0.9 and 0.1. VesselFM is trained on a single V100 GPU (32GB) with a batch size of 8 until convergence.
The learning rate is set to $\text{10}^\text{-4}$. We utilize linear warm-up and a learning rate decay to $\text{10}^\text{-5}$. During training, we sample classes near uniformly from $\mathcal{D}_\text{real}$ and also from $\mathcal{D}_\text{flow}$.
We solely apply data augmentation to samples from $\mathcal{D}_\text{real}$. To optimize training efficiency, we perform data augmentation offline. Specifically, we apply, after extracting $\text{128}^{\text{3}}$ patches, random flipping and rotation (angle in [0$^\circ$, 10$^\circ$]) along all axes followed by random elastic deformation ($\sigma$ in [10, 20] and magnitude in [100, 500]) and random zooming (factor in [0.9, 1.3]). 

\paragraph{\textit{One}- and \textit{few}-shot fine-tuning.}
In the \textit{one}- and \textit{few}-shot setting, we fine-tune vesselFM using a similar setup with a learning rate of $\text{10}^\text{-5}$. We train vesselFM until convergence, selecting the checkpoint with the best Dice score on the respective validation volume.
We apply lightweight data augmentations on the fly: random zooming (factor in [1, 1.3]), random shearing (shearing factors in [0, 0.4]), random flipping, random Gaussian smoothing ($\sigma$ in [0, 0.5]), random Gaussian noise ($\mu$ of 0.3, $\sigma$ in [0, 0.01]), and random histogram shifting (number of points in [5, 10]).

Following our general procedure, we fine-tune all baselines until convergence and select the checkpoint with the best Dice score on the validation volume for testing.
\textit{tUbeNet}~\cite{holroyd2023tube}:
We fine-tune tUbeNet using our training scheme described above, solely adapting the patch size (tUbeNet operates on patches of shape $\text{64}^\text{3}$) and employing its linear learning rate decay. We further apply the same lightweight data augmentations used for fine-tuning vesselFM.
\textit{VISTA3D}~\cite{he2024vista3d}:
We fine-tune VISTA3D with the script provided by the authors.
As VISTA3D predicts 127 classes, we default to the only class representing blood vessels, the hepatic vessel class. Given that VISTA3D is designed specifically for CT images, we replace their interval-based intensity scaling scheme with a percentile-based scaling scheme (2 and 98 percentiles) and omit their resampling transformation. Further augmentations are left unchanged. During inference, we use the default "auto + point" configuration, which has been shown to yield the best results.
\textit{SAM-Med3D}~\cite{wang2024sammed3d}:
We adopt SAM-Med3D's training and inference pipeline without major changes. We utilize their default setting, providing one query point during training and five during inference. Since their data augmentation pipeline closely resembles vesselFM's, we retain SAM-Med3D's without alterations.
\textit{MedSAM-2}~\cite{zhu2024medical}:
As MedSAM-2 is trained on images of size 1024, we resample patches used for fine-tuning. Other than that, we keep their original setup, which fine-tunes the mask decoder of the SAM 2 model and the memory layer, unchanged. We use the default configuration of one query point in every second slice for both training and inference.

\section{VesselFM From Scratch}\label{suppl:scratch}
'VesselFM (from scratch)' (see Table~\ref{tab:quantitative_results}) demonstrates a relatively strong performance compared to other baselines and even outperforms them on the OCTA and BvEM datasets in some metrics. We attribute this to their unique imaging artifacts and intensity patterns, which, unlike those in CT and MRA, have not been observed by any baseline during pre-training. Note that the human retinal OCTA sample present in tUbeNet's training dataset exhibits poor label quality and differs significantly from the characteristics~\cite{wittmann2024simulation} of the murine cerebral OCTA images used in our experiments.
Further, architectural biases (SAM-Med3D and MedSAM-2 rely on large Transformers) and mismatches in already learned representations (VISTA3D is exclusively trained on CT images; tUbeNet is exclusively trained on images with stark contrast to background tissues) may impede few-shot fine-tunability of baselines. This, together with 'vesselFM (from scratch)' benefiting from our well-evaluated UNet architecture, explains its relatively strong performance.

\section{More Detailed Ablation of Data Sources}\label{suppl:source_abl}
A more detailed ablation study of the effect of supplementing $\mathcal{D}_\text{real}$ with $\mathcal{D}_\text{drand}$ and $\mathcal{D}_\text{flow}$, covering all evaluation datasets and tasks, is shown in Table~\ref{tab:suppl_ablation_data_source}. We find that the combination of all of our three proposed data sources generally yields the best segmentation performance.

To further substantiate this hypothesis, we investigate how the performance of vesselFM scales with the amount of available training data for \textit{zero}-shot segmentation. To this end, we progressively augment $\mathcal{D}_\text{real}$ with 10\%, 50\%, and finally 100\% of the data from our two synthetic data sources, $\mathcal{D}_\text{drand}$ and $\mathcal{D}_\text{flow}$. We scale weights assigned to data sources accordingly. Similar to Table~\ref{tab:suppl_ablation_data_source}, we first augment $\mathcal{D}_\text{real}$ with $\mathcal{D}_\text{drand}$, followed by $\mathcal{D}_\text{flow}$. Our findings are demonstrated in Fig.~\ref{fig:suppl_amount_of_training_data}. We observe a significant performance increase introducing $\mathcal{D}_\text{drand}$, which flattens as it approaches 100\%. On the SMILE-UHURA and BvEM datasets, performance additionally spikes after introducing $\mathcal{D}_\text{flow}$, while performance on OCTA and MSD8 stagnates. 
Averaged across all four evaluation datasets (see Fig.~\ref{fig:suppl_amount_of_training_data}, average), we find that the additional diversity introduced by $\mathcal{D}_\text{flow}$ proves to be consistently beneficial for segmentation performance.

\begin{figure}[t]
\centerline{\includegraphics[width=\linewidth]{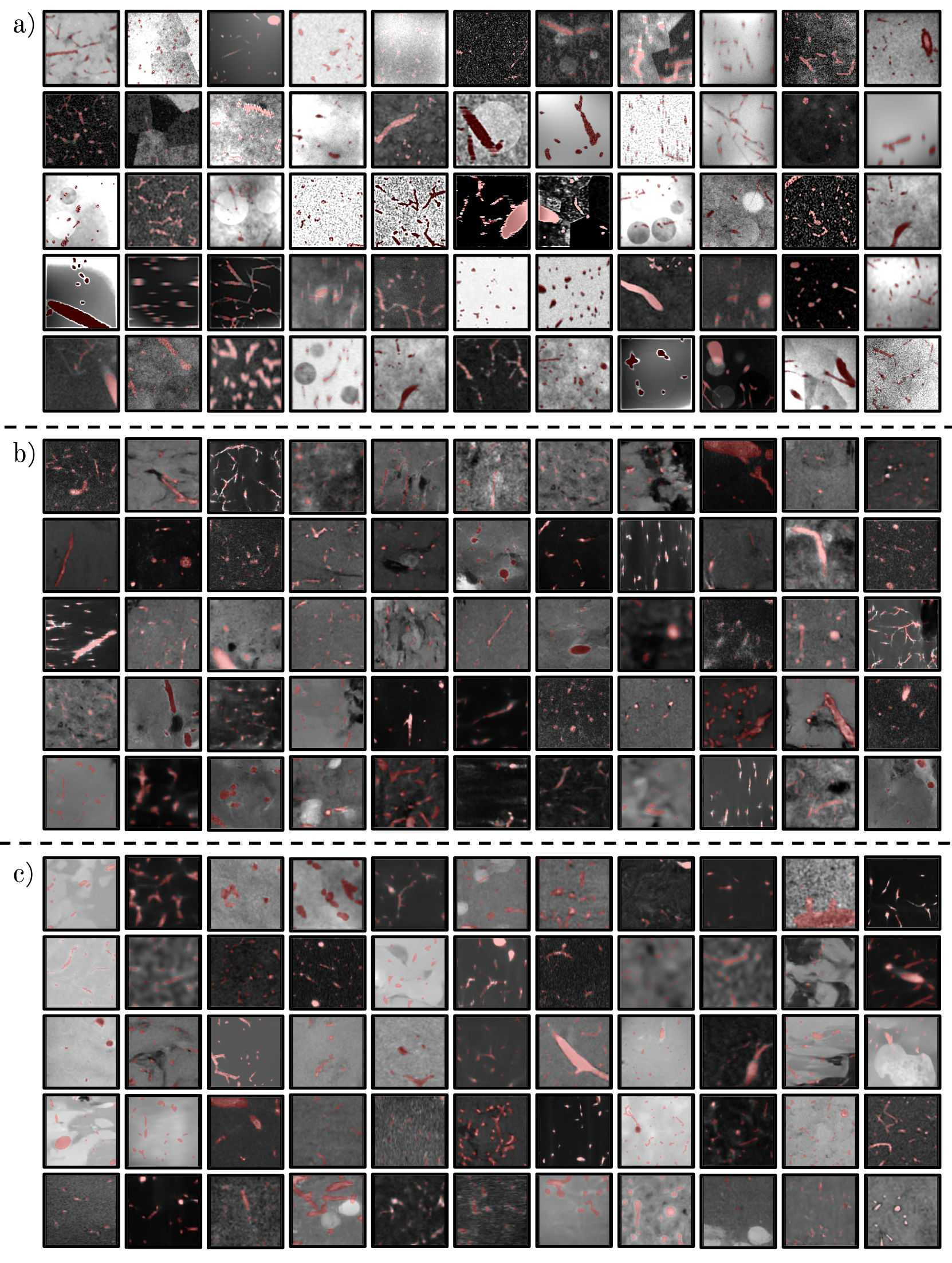}}
\caption{
More slices of exemplary domain randomized images (a), images sampled from our flow matching-based generative model $\mathcal{F}$ (b), and images sampled from Med-DDPM (c). Masks are shown in translucent red.
}
\label{fig:suppl_more_samples}
\end{figure}

\section{VesselFM's Utility in (Pre-)Clinical Settings}\label{suppl:clinical}
In (pre-)clinical settings, high-quality annotations for emerging imaging technologies and novel, unstudied structures of interest are often not immediately available. Therefore, clinicians and researchers typically create voxel-level annotations from scratch through labor-intensive manual labeling to train supervised segmentation algorithms, a necessity for automated, large-scale, and accurate analysis. Voxel-level annotations, however, can be far more efficiently obtained via automated pre-segmentation and iterative label refinement. This process is often referred to as bootstrapping. With specialist models failing to bridge domain gaps, vesselFM's exceptional \textit{zero}-shot generalization and fine-tunability render it ideal for such applications.

\begin{figure}[t]
\centerline{\includegraphics[width=\linewidth]{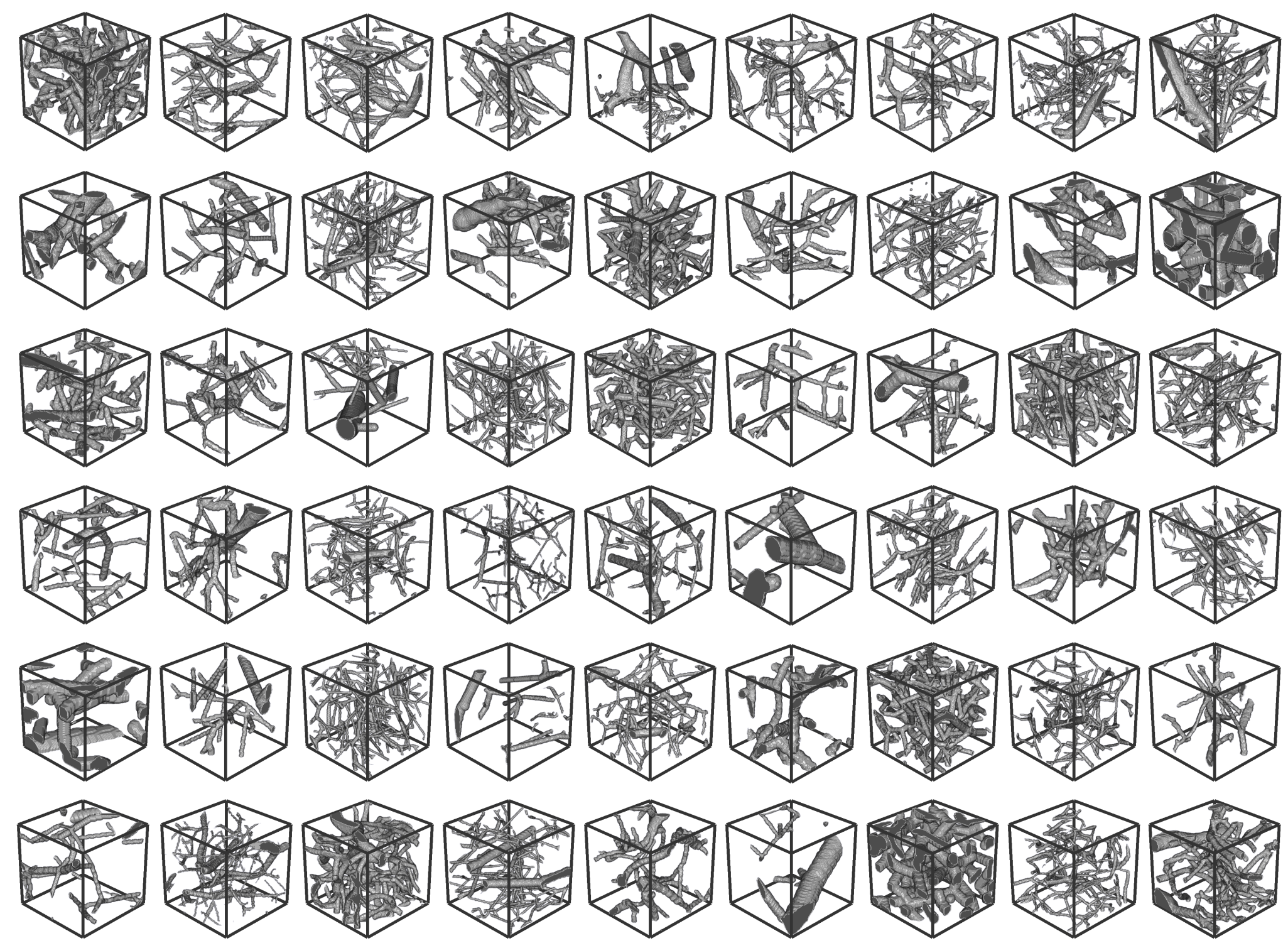}}
\caption{More exemplary synthetic masks $\mathcal{M}_\text{syn}$, generated by our proposed domain randomized generative pipeline (see Fig.~\ref{fig:method_drand}a).}
\label{fig:suppl_masks}
\end{figure}

\begin{table*}[ht]
\centering
\scriptsize
\caption{Comparison to specialist models trained on individual blood vessel segmentation datasets.}
\label{tab:specialist_models}
\begin{tabular}{c|l| c c| c c| c c| c c} 
\toprule
\multirow{2}{*}{Task} & \multirow{2}{*}{Model} &
\multicolumn{2}{c|}{OCTA~\cite{wittmann2024simulation,glandorf2024bessel}} & \multicolumn{2}{c|}{BvEM~\cite{wan2024trisam}} & \multicolumn{2}{c|}{SMILE-UHURA~\cite{chatterjee2024smile}}& \multicolumn{2}{c}{MSD8~\cite{antonelli2022medical}}\\
&& $\text{Dice}\uparrow$ & $\text{clDice}\uparrow$ & $\text{Dice}\uparrow$ & $\text{clDice}\uparrow$ & $\text{Dice}\uparrow$ & $\text{clDice}\uparrow$ & $\text{Dice}\uparrow$ & $\text{clDice}\uparrow$\\
\midrule
\multirow{4}{*}{\rotatebox[origin=c]{90}{\textit{zero-shot}}}   
& specialist model - TopCoW~\cite{yang2023benchmarking} & 39.14 & 24.03 & 55.09 & 52.42 & 38.71 & 34.35 & 4.18 & 3.45\\
& specialist model - VesSAP~\cite{todorov2020machine} & 33.31 & 60.69 & 50.89 & 32.75 & 16.27 & 25.63 & 9.40 & 15.19\\
& specialist model - CSD~\cite{chen2022attention,ixidataset} & 21.92 & 42.98 & 0.04 & 0.01 & 59.17 & 53.07 & 0.44 & -\\
\cmidrule{2-10}
& vesselFM (ours) & \textbf{46.94} & \textbf{67.07} & \textbf{67.49} & \textbf{62.04} & \textbf{74.66} & \textbf{75.27} & \textbf{29.69} & \textbf{36.14}\\
\midrule
\multirow{4}{*}{\rotatebox[origin=c]{90}{\textit{one-shot}}}   
& specialist model - TopCoW~\cite{yang2023benchmarking} & 64.80 & 74.39 & 71.53 & 78.65 & 36.52 & 35.66 & 17.61 & -\\
& specialist model - VesSAP~\cite{todorov2020machine} & 69.26 & 76.67 & 72.31 & 75.94 & 27.40 & 30.50 & 28.94 & 35.40\\
& specialist model - CSD~\cite{chen2022attention,ixidataset} & 64.62 & 77.56 & 68.64 & 78.96 & 75.96 & 78.20 & 16.09 & -\\
\cmidrule{2-10}
& vesselFM (ours) & \textbf{72.10} & \textbf{83.73} & \textbf{78.27} & \textbf{79.91} & \textbf{76.43} & \textbf{78.36} & \textbf{36.88} & \textbf{48.65}\\
\midrule
\multirow{4}{*}{\rotatebox[origin=c]{90}{\textit{few-shot}}}   
& specialist model - TopCoW~\cite{yang2023benchmarking} & 70.28 & 76.79 & 65.89 & 77.76 & 52.97 & 47.54 & 26.25 & 37.16\\
& specialist model - VesSAP~\cite{todorov2020machine} & 70.27 & 77.14 & 69.36 & 78.40 & 46.83 & 44.31 & 36.27 & 44.74\\
& specialist model - CSD~\cite{chen2022attention,ixidataset} & 71.11 & 79.36 & 59.94 & 78.61 & 78.20 & 78.59 & 34.28 & 47.57\\
\cmidrule{2-10}
& vesselFM (ours) & \textbf{75.70} & \textbf{84.03} & \textbf{78.11} & \textbf{84.54} & \textbf{78.77} & \textbf{79.37} & \textbf{45.04} & \textbf{57.25}\\
\bottomrule
\end{tabular}
\end{table*}

\section{More Samples From \texorpdfstring{$\mathcal{D}_\text{drand}$}{} and \texorpdfstring{$\mathcal{D}_\text{flow}$}{}}
We present additional samples from $\mathcal{D}_\text{drand}$ and $\mathcal{D}_\text{flow}$ in Fig.~\ref{fig:suppl_more_samples}a and b, respectively. In conclusion, one can state that our domain randomized generative pipeline produces a wide variety of image-mask pairs with highly diverse fore- and background geometries and textures, while images sampled from our flow matching-based generative model $\mathcal{F}$ exhibit intensity patterns closely mimicking those of real images in $\mathcal{D}_\text{real}$. Please note that Fig.~\ref{fig:suppl_more_samples}c depicts images sampled from the Med-DDPM baseline for comparison.

\section{More Masks \texorpdfstring{$\mathcal{M}_\text{syn}$}{}}
To further showcase the wide variety of synthetic masks $\mathcal{M}_\text{syn}$ produced by our proposed domain randomized generative pipeline, we present a comprehensive selection in Fig.~\ref{fig:suppl_masks}. Masks contained in $\mathcal{M}_\text{syn}$ encompass a broad range of realistic vascular patterns, capturing variations in blood vessel scale, density, curvature, and tortuosity.

\section{Additional Qualitative \textit{Zero}-Shot Results}
\begin{figure*}[t]
\centerline{\includegraphics[width=0.90\linewidth]{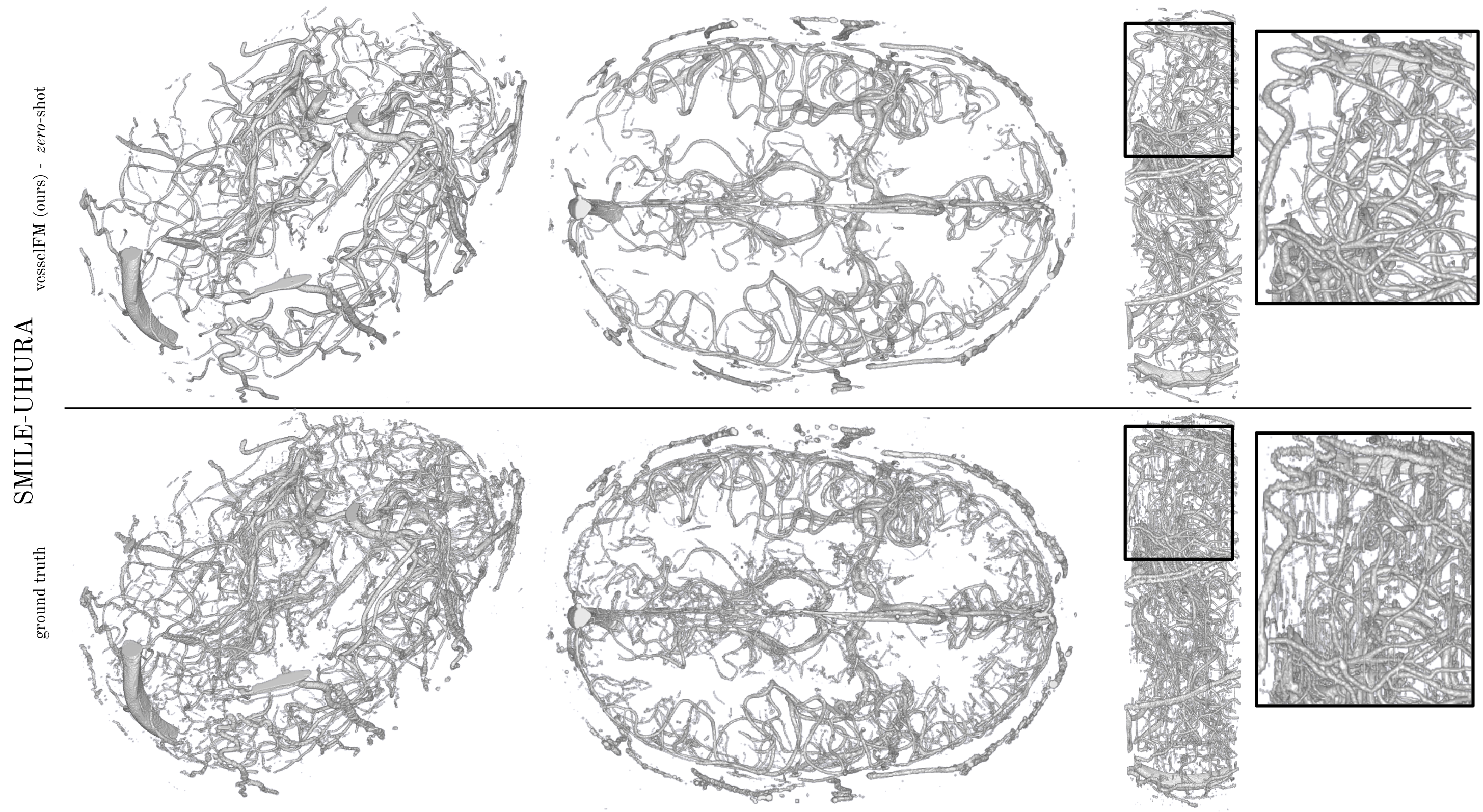}}
\caption{
Qualitative results achieved on an exemplary test sample from the SMILE-UHURA dataset~\cite{chatterjee2024smile}. We compare vesselFM's prediction in the \textit{zero}-shot setting (top row) to the ground truth label contained in the SMILE-UHURA dataset (bottom row). VesselFM delivers remarkable results free of artifacts and accurately maintains the tubular appearance of blood vessels (see black box).
}
\label{fig:suppl_qual_res_smile}
\end{figure*}

\begin{figure*}[t]
\centerline{\includegraphics[width=0.9\linewidth]{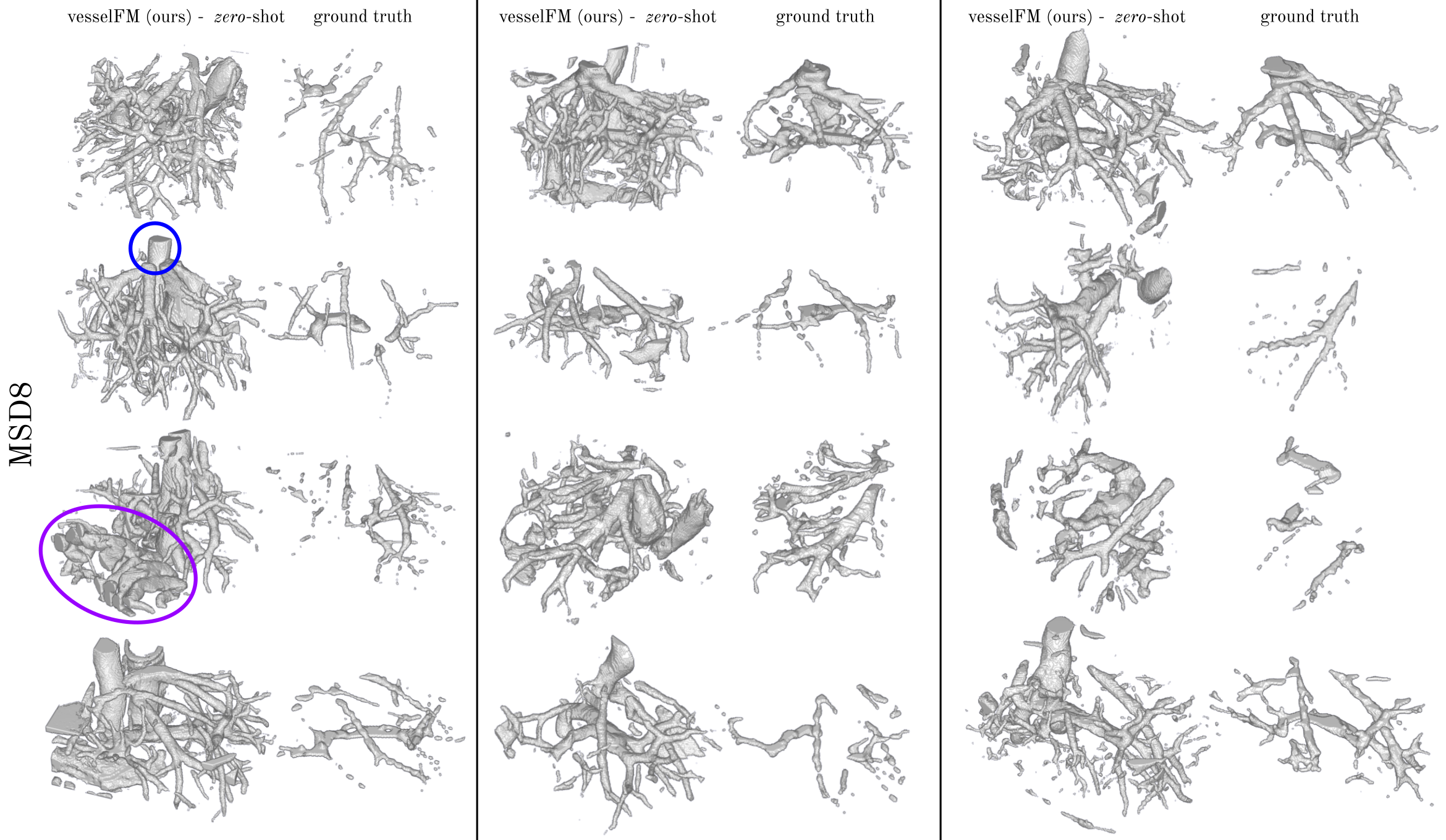}}
\caption{
Qualitative results achieved on multiple test samples from the MSD8 dataset~\cite{antonelli2022medical}. We compare vesselFM's predictions in the \textit{zero}-shot setting to ground truth labels for the task of hepatic vessel segmentation contained in the MSD8 dataset. VesselFM accurately segments all blood vessels (\eg, aorta (marked in blue) and other major components of the systemic arterial circulation) and even other tubular structures (\eg, the colon (marked in purple) and parts of the rib cage) present in CT scans.
}
\label{fig:suppl_qual_res_msd8}
\end{figure*}

\begin{figure*}[t]
\centerline{\includegraphics[width=0.9\linewidth]{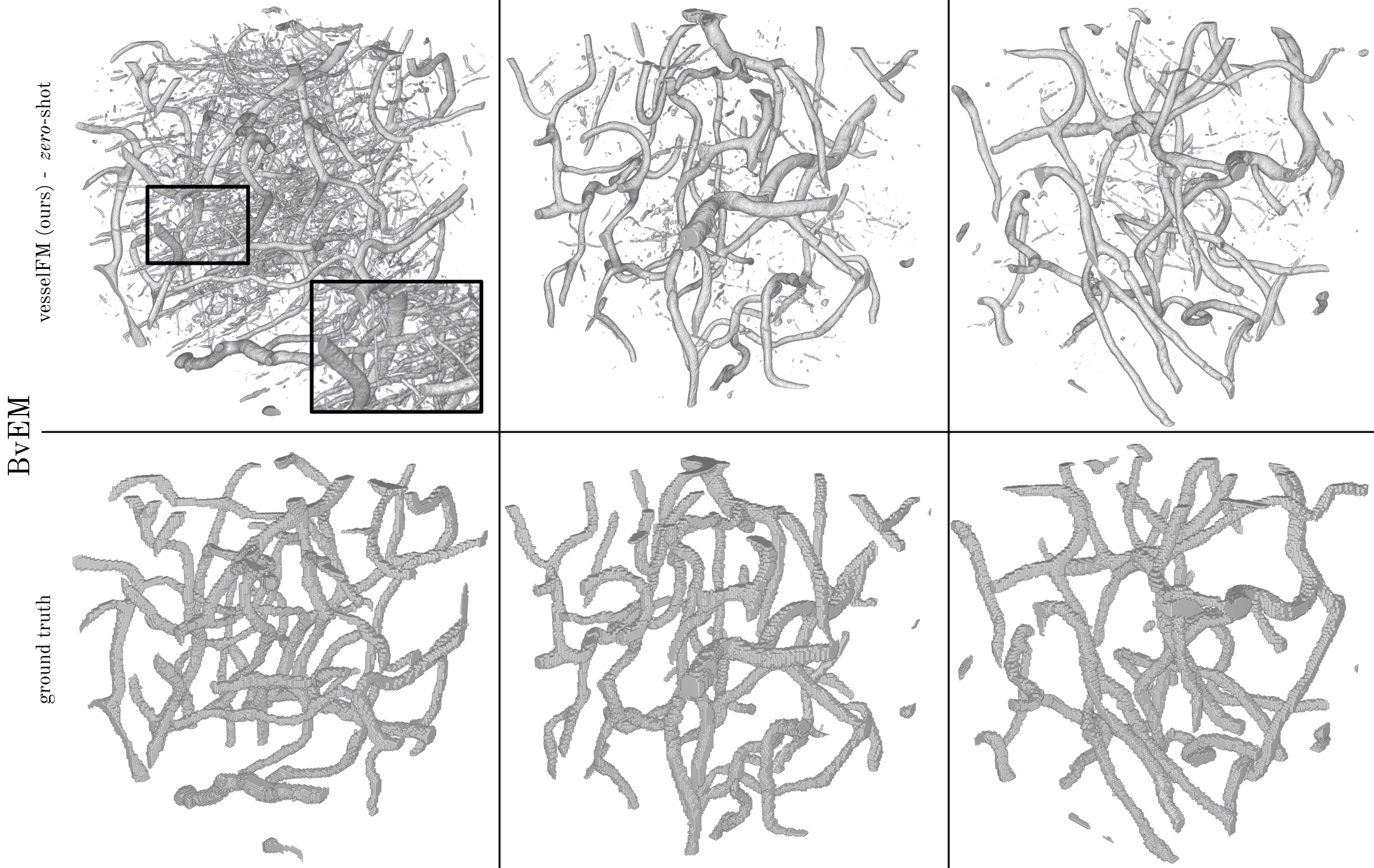}}
\caption{
Qualitative results achieved on the three test volumes extracted from the BvEM dataset~\cite{wan2024trisam}. We compare vesselFM's predictions in the \textit{zero}-shot setting (top row) to ground truth labels contained in the BvEM dataset (bottom row). We find that vesselFM segments murine cortical vasculature contained in volume electron microscopy (vEM) images very precisely. Interestingly, vesselFM segments not only blood vessels but also tubular-appearing axons and even dendrites of pyramidal cells (see black box) visible in vEM images~\cite{microns2021functional}.
}
\label{fig:suppl_qual_res_bvem}
\end{figure*}

\begin{figure*}[t]
\centerline{\includegraphics[width=0.9\linewidth]{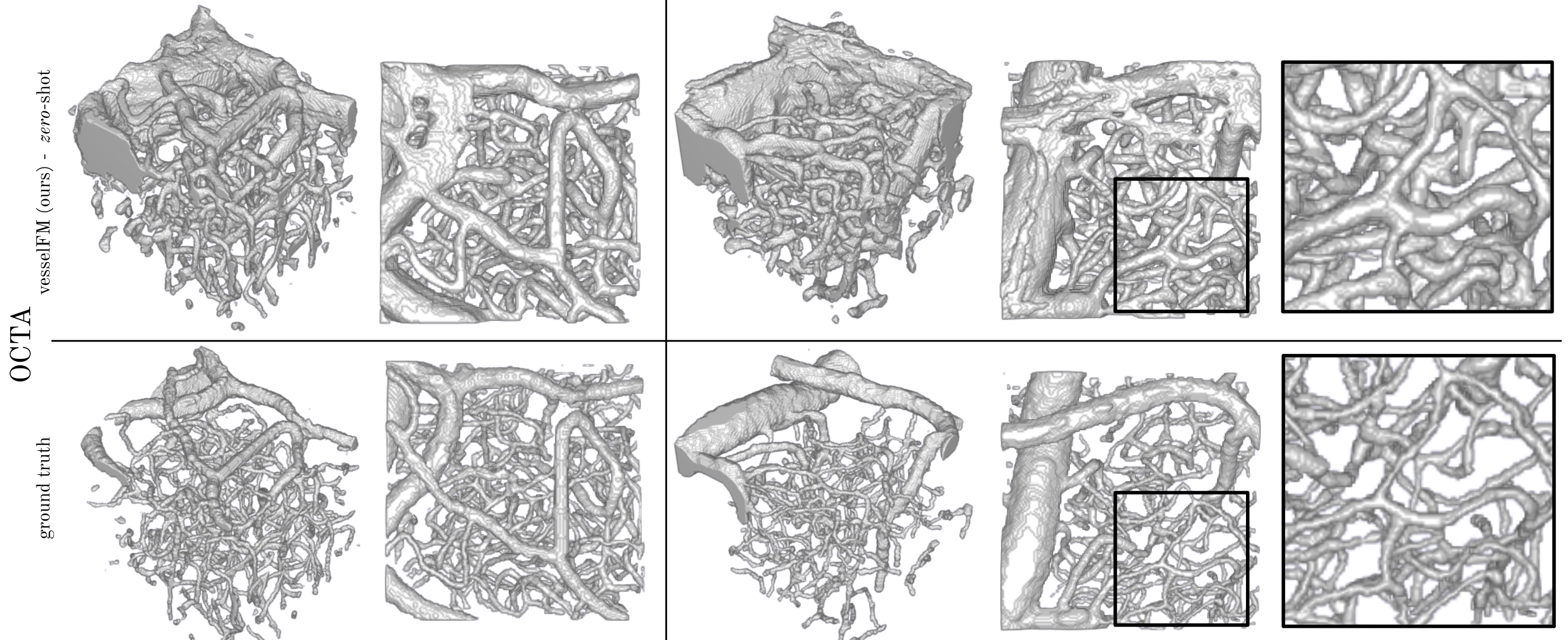}}
\caption{
Qualitative results achieved on two test samples from the OCTA dataset~\cite{wittmann2024simulation,glandorf2024bessel}. We compare vesselFM's predictions in the \textit{zero}-shot setting (top row) to ground truth labels contained in the OCTA dataset (bottom row). Although OCTA images are known for being plagued by dominant imaging artifacts~\cite{li2022blood,hormel2021artifacts,zhu2020visibility}, vesselFM still manages to segment densely connected vasculature (see black box).
}
\label{fig:suppl_qual_res_octa}
\end{figure*}

To emphasize the exceptional \textit{zero}-shot generalization of vesselFM, we present additional qualitative results achieved on all four evaluation datasets (see~\cref{fig:suppl_qual_res_smile,fig:suppl_qual_res_msd8,fig:suppl_qual_res_bvem,fig:suppl_qual_res_octa}). Our findings demonstrate that vesselFM segments blood vessels very accurately across all evaluation datasets. Interestingly, vesselFM also segments tubular-appearing structures beyond blood vessels (\eg, axons (see Fig.~\ref{fig:suppl_qual_res_bvem}) or parts of the colon (see Fig.~\ref{fig:suppl_qual_res_msd8})). This highlights vesselFM's strong inductive bias towards tubular shapes. 

By segmenting all tubular structures in the volume, vesselFM segments structures, which are, at least to some degree, not annotated in ground truth labels (\eg, aorta or other components of the systemic arterial circulation in MSD8 (see Fig.~\ref{fig:suppl_qual_res_msd8})). We argue that this may artificially deflate vesselFM's quantitative results reported for the \textit{zero}-shot task in Table~\ref{tab:quantitative_results}.

\section{Additional Flow Matching Trajectories}
Fig.~\ref{fig:suppl_trajectories} presents additional sampled flow matching trajectories, similar to Fig.~\ref{fig:method_fm}. Specifically, we visualize the mapping from $x_{0} \sim \mathcal{N}(0, I)$ to samples $x_{1}$ of exemplary classes indicated in red.

\section{Statistical Analysis of Results}
We conducted a statistical analysis of the quantitative results reported in the main manuscript using paired t-tests. To this end, we compared vesselFM's base configuration to the respective runner-up. We found all results to be statistically significant ($p$ $<$ 0.05), except for the Dice score in the ablation on class conditioning (Table~\ref{tab:ablations_flow}, 4th row; $p$ = 0.081). The clDice score in the ablation on class conditioning, however, remains significant ($p$ = 0.036). Notably, values reported in Table~\ref{tab:ablations_arch} are statistically significant ($p$ = 0.019 for Dice; $p$ = 2.39 $\cdot$ $\text{10}^{-\text{5}}$ for clDice).

\section{Computational Resources}
VesselFM comprises 31,418,977 parameters and requires minimal computational resources compared to other foundation models for 3D image segmentation. Processing a volume of shape $\text{128}^\text{3}$ takes 335.7 ms on a T4 GPU, 95.8 ms on a V100 GPU, 30.6 ms on an A100 GPU, and 3.3 s on an AMD Epyc 7702 CPU, with a VRAM consumption of approximately 4.21~GB.

\section{Comparison to Specialist Models}
We further compare vesselFM to specialist models trained on individual blood vessel segmentation datasets, as shown in Table~\ref{tab:specialist_models}. As expected, vesselFM not only consistently outperforms state-of-the-art foundation models for medical image segmentation, but also specialist models, even when pre-trained on exactly the same imaging modality (CSD and SMILE-UHURA both contain MRA images).

\clearpage

\end{document}